\documentclass[12pt]{article}
\usepackage[english]{babel}

\usepackage[nottoc,notlot,notlof]{tocbibind}
\usepackage{amssymb,mathrsfs,amsfonts}
\usepackage{amsmath, amsthm, xpatch, dsfont,ushort,sgamevar, float, framed, multirow, array, makecell, comment, url, graphicx, subcaption, mathtools, fnpct, setspace, pifont, bm, bbm,accents}
\usepackage{nicefrac, xfrac}
\usepackage{epstopdf}
\usepackage[normalem]{ulem}

\usepackage{lmodern}
\usepackage[final, babel, activate={true,nocompatibility}]{microtype}
\DeclarePairedDelimiterX{\infdivx}[2]{(}{)}{%
#1\;\delimsize\|\;#2%
}

\usepackage{lipsum}
\usepackage[utf8]{inputenc} 
\usepackage[T1]{fontenc} 
\usepackage{titlesec}
\usepackage{thmtools}

\usepackage[round,comma,authoryear,longnamesfirst,sort&compress]{natbib}
\usepackage[bottom, multiple]{footmisc}
\usepackage[top=1.25in, bottom=1.25in, left=1.25in, right=1.25in]{geometry}
\usepackage{xcolor}
\definecolor{navyblue}{rgb}{0.0, 0.13, 0.5}
\definecolor{amaranth}{rgb}{0.9, 0.17, 0.31}
\definecolor{brightpink}{rgb}{1.0, 0.0, 0.5}
\definecolor{cgreen}{rgb}{0.0, 0.42, 0.24}
\definecolor{cobalt}{rgb}{0.0, 0.28, 0.67}
\definecolor{coquelicot}{rgb}{1.0, 0.22, 0.0}
\definecolor{deepcarrotorange}{rgb}{0.91, 0.41, 0.17}
\definecolor{portlandorange}{rgb}{1.0, 0.35, 0.21}
\definecolor{burntorange}{rgb}{0.8, 0.33, 0.0}
\definecolor{dukeblue}{rgb}{0.0, 0.0, 0.61}
\definecolor{brickred}{rgb}{0.8, 0.25, 0.33}
\definecolor{brown(web)}{rgb}{0.65, 0.16, 0.16}
\definecolor{ceruleanblue}{rgb}{0.16, 0.32, 0.75}

\usepackage{hyperref}
\hypersetup{
colorlinks=true,
linkcolor=dukeblue,
filecolor=dukeblue,
citecolor=dukeblue,
urlcolor=dukeblue,
}
\usepackage[capitalize,noabbrev,nameinlink]{cleveref}

\crefformat{onlineappendix}{#2Supplementary Appendix~#1#3}
\Crefformat{onlineappendix}{#2Supplementary Appendix~#1#3}

\DeclareMathAlphabet{\mathsf}{OT1}{\sfdefault}{m}{n}
\SetMathAlphabet{\mathsf}{bold}{OT1}{\sfdefault}{b}{n}

\usepackage{enumitem}

\newcommand{\argmax}{\arg\!\max}

\renewcommand\thmcontinues[1]{Continued}

\newtheorem{claim*}{Claim}[]

\newcommand*\diff{\mathop{}\!\mathrm{d}}
\usepackage{tocloft}

\newtheorem{prop*}{Proposition}[]
\newtheorem{bprop*}{Proposition}
\newtheorem{bthm*}{Theorem}
\newtheorem{thm*}{Theorem}[]

\newtheorem{ax*}{Axiom}[]
\newtheorem{lem*}{Lemma}[]
\newtheorem{rem*}{Remark}[]
\newtheorem{asp*}{Assumption}[]
\newtheorem{cor*}{Corollary}[]
\newtheorem{pty*}{Property}[]
\newtheorem{def*}{Definition}[]
\newtheorem{cond*}{Condition}[]

\newtheorem*{propnonumber*}{Proposition}

\usepackage{caption}
\usepackage[labelfont=bf]{caption}

\usepackage{tikz}
\usepackage{pgfplots}
\usetikzlibrary{automata,positioning}
\usetikzlibrary{arrows,intersections}

\newtheorem{examplex}{Example}
\newenvironment{eg*}
{\pushQED{\qed}\examplex}
{\popQED\endexamplex}

\newcommand{\ra}{\rightarrow}

\newcommand{\bE}{\mathbf{E}}
\newcommand{\bR}{\mathbf{R}}
\newcommand{\bN}{\mathbb{N}}

\newcommand{\gb}{\beta}
\newcommand{\gd}{\delta}
\newcommand{\gs}{\sigma}

\newcommand{\egg}{e.g.,~}
\newcommand{\ie}{i.e.,~}

\usepackage[titletoc,title]{appendix}

\renewcommand{\texttt}[1]{%
\begingroup
\ttfamily
\begingroup\lccode`~=`/\lowercase{\endgroup\def~}{/\discretionary{}{}{}}%
\begingroup\lccode`~=`[\lowercase{\endgroup\def~}{[\discretionary{}{}{}}%
\begingroup\lccode`~=`.\lowercase{\endgroup\def~}{.\discretionary{}{}{}}%
\catcode`/=\active\catcode`[=\active\catcode`.=\active
\scantokens{#1\noexpand}%
\endgroup
}

\makeatletter
\renewenvironment{proof}[1]
[\proofname]{\par\pushQED{\qed}\normalfont%
\topsep6\p@\@plus6\p@\relax
\trivlist\item[\hskip\labelsep\bfseries#1\@addpunct{.}]%
\ignorespaces}{%
\popQED\endtrivlist\@endpefalse
}
\makeatother

\pgfplotsset{compat=1.14}
\usetikzlibrary{automata,arrows.meta,positioning,shapes.geometric}

\usepackage{placeins}
\usepackage{needspace}

\makeatletter
\newcommand{\vast}{\bBigg@{4}}
\newcommand{\Vast}{\bBigg@{6}}
\makeatother

\makeatletter
\AddToHook{cmd/appendix/before}{\def\cref@section@alias{appendix}\def\cref@subsection@alias{appendix}}
\makeatother

\begin{document}
\title{Task Assignment with Interdependent Incentives\thanks{We thank Parimal Bag, In\'acio B\'o, Yeon-Koo Che, Lawrence Choo, Angus Chu, Shanglyu Deng, Mehmet Ekmekci, Hanming Fang, Sam Jindani, Fei Li, Jiangtao Li, Jin Li, Marina Halac, George Mailath, \href{https://refine.ink/}{Refine.ink}, Larry Samuelson, Bal\'azs Szentes, Satoru Takahashi, Tat-How Teh, Tim Worrall, Zhaoneng Yuan, and many seminar audiences for helpful comments.}}
\author{Yonghang Ji\thanks{University of Macau, \href{mailto:yonghangji@outlook.com}{\texttt{yonghangji@outlook.com}}.}
\and
Allen Vong\thanks{National University of Singapore, \href{mailto:allenv@nus.edu.sg}{\texttt{allenv@nus.edu.sg}}.}}
\date{\today}

\maketitle

\begin{abstract}{
We study a manager's repeated assignment of a burden among a group of homogeneous workers under imperfect public monitoring of workers' effort. Their effort incentives are interdependent because assignment is exclusive. We develop recursive bounds on workers' assignment frequencies to address this interdependence and characterize the manager's optimum among equilibria inducing effort whenever assignment occurs. We also explicitly construct an equilibrium attaining this optimum. Whenever some effortful assignment is sustainable, this equilibrium retains the assignee after failure and relieves him after success until each other worker has produced a success and a vacancy phase has elapsed. Our results extend to assignment of a desirable task.
}
\end{abstract}

\thispagestyle{empty}
\pagebreak
\setcounter{page}{1}

\section{Introduction}
\label{sec:introduction}

Managers routinely assign workers to particular responsibilities. This discretion not only allocates production but is also a powerful incentive instrument: relief from an onerous duty or access to a prominent project can reward performance and motivate workers to exert effort. Effort incentives driven by assignment are distinctive because they are interdependent: assigning one worker a burden, for example, spares others from it. How should organizations structure these incentives?

To address this question, we study a long-run relationship in which assignment is the only incentive instrument. Each period, a manager either leaves a task vacant or assigns it to one of several ex ante identical workers; in our baseline model, the task is undesirable for the workers. The assignee privately chooses whether to exert effort or shirk, generating a noisy public outcome of success or failure. Effort is costly to the worker but increases the probability of success, which benefits the manager. The manager cannot commit and may condition future assignments on the entire history of assignments and outcomes. We call her strategy an assignment rule. We ask how assignments should be structured to induce effort whenever a worker is assigned on path, and call an equilibrium with this property a no-shirking equilibrium.

Our setting and focus on no-shirking equilibria speak to high-stakes responsibilities assigned to one of several similarly qualified workers. Examples include anesthesia coverage during surgery, an air-traffic-control position, a maritime or military watch, and cyber incident command. In each case, diligence is crucial, and operational outcomes---such as whether a patient remains stable, aircraft remain safely separated, a hazard is detected during a watch, or a cyber incident is contained---provide noisy signals of effort.

Our main result characterizes the manager's maximum payoff among all no-shirking equilibria and shows that a ``first-order-rotation'' assignment rule attains it whenever a nontrivial no-shirking equilibrium---one in which some effortful assignment occurs---exists. Under this rule, failure retains the current assignee, whereas success relieves him until every other worker has been assigned and succeeded once, and a vacancy phase has elapsed. The vacancy-phase length increases with the severity of the moral hazard problem, ranging from zero---corresponding to uninterrupted effortful operation---to infinity---corresponding to complete shutdown after one rotation phase. This rule is first-order in the sense that each current assignee's retention or relief depends only on his current outcome. Since this rule rewards success with respite from an undesirable task, it intuitively sustains effort when moral hazard is mild. Our result is substantially stronger: the constructed first-order rotation attains the manager's optimum among arbitrary history-contingent rules.

The challenge in characterizing the manager's optimum lies in preserving all workers' effort incentives while allocating the task. With at least three workers, relieving the current assignee by reassigning the task requires choosing among multiple replacements. This choice affects the distribution of future assignments among the workers and, in turn, their effort incentives. This interdependence corresponds to a familiar difficulty in repeated principal-multi-agent games without transfers: characterizing the geometry of the equilibrium-payoff set. \citet{de2021selecting} illustrate this difficulty in their adverse-selection setting, even when attention is restricted to equilibria attaining the principal's first-best. With two agents, the interval structure of the agents' principal-first-best equilibrium-payoff set makes the extreme continuation payoffs and thus the scope of incentive provision tractable, permitting an explicit characterization of the first-best region---the region of parameters for which a principal-first-best equilibrium exists; see also \citet{athey2001optimal}. With three or more agents, this approach does not extend directly because the payoff-set geometry is unknown, and they do not characterize the first-best region.\footnote{The self-generation method of \citet{abreu1990toward} characterizes equilibrium-payoff sets recursively but does not by itself provide a closed-form solution.}

Our methodological contribution is a recursive approach that bypasses the payoff-set problem in our moral-hazard setting. We explicitly characterize, for any workforce size, both the first-best parameter region and the larger parameter region in which a nontrivial no-shirking equilibrium exists. For each group size, we bound the largest total average assignment frequency attainable by any group of that size in any no-shirking equilibrium and show that these bounds lead to a recursion that upper-bounds the manager's payoff. We then construct a first-order rotation rule that attains this upper bound and is therefore optimal whenever a nontrivial no-shirking equilibrium exists, whether or not the manager's first-best is attainable.

Our characterization has two substantive implications. First, our result shows how staffing, working conditions, and monitoring shape the effectiveness of assignment incentives. A larger workforce makes the manager's first-best---uninterrupted effortful operation---easier to sustain. This effect contrasts with tournaments in which adding contestants can dilute effort incentives (\citealp{nalebuff1983prizes}; \citealp{taylor1995digging}). We further show that whenever first-order rotation sustains a no-shirking equilibrium with finite vacancy phases, a sufficiently large workforce allows the manager to attain her first-best. First-order rotation also differs from tournaments in that it does not use relative-performance comparisons: an assignee's transition depends only on his own current outcome. More valuable relief, a higher success probability under effort, and a lower success probability under shirking also expand the first-best region. These comparative statics matter for organizational design when interruptions are prohibitively costly or infeasible, as in anesthesia during surgery or air-traffic control. We further show that the same forces extend beyond the first-best region and shape the manager's optimal payoff.

Second, our comparison with a random-reselection assignment rule inspired by \citet{de2021selecting} identifies an incentive benefit of a persistent rotation order. Under first-order rotation, the initial ordering may be drawn at random, but once realized it persists: workers with identical outcome records can face different future assignment frequencies solely because of their queue positions. Random reselection can instead return a successful worker to the task before every other worker has served since his success, shortening the respite that success earns. We show that with three or more workers, first-order rotation expands the first-best region and can strictly raise the manager's payoff relative to the random-reselection rule; with two workers, both classes attain the same optimal payoff. This finding provides a new rationale for nonmonetary discrimination in organizations: whereas unequal workload may be attributed to differences in worker characteristics or past performance, we show that it can arise endogenously even among ex ante identical workers with identical performance records as an instrument of dynamic incentive provision.

Our conclusions extend in several directions that we discuss in \cref{sec:extensions}. In particular, they extend to a desirable rather than undesirable task, with a reversed direction of first-order rotation: success leads to retention, whereas failure leads to removal---reflecting that assignment is then a reward rather than a punishment.

\paragraph{Related literature.} We contribute to the literature on dynamic principal-multi-agent games. The closest paper is \citet{de2021selecting}, who also study repeated assignment among ex ante identical agents, with binary outcomes, no transfers, and no commitment. Their friction is adverse selection: agents privately learn whether they are qualified but want the task regardless, whereas the principal wants to select a qualified agent. With two agents, they characterize the parameter region over which the principal can attain her first-best and construct a rule that attains her first-best throughout this region. With more agents, they leave the characterization of the first-best region open and instead study a rule that treats all but one designated agent symmetrically in each period, showing that no rule within their class of hierarchical strategy profiles dominates it. We instead study moral hazard and develop a method to characterize the manager's maximum payoff among all no-shirking equilibria for any workforce size and any parameters, including but not limited to those that attain the manager's first-best. Our optimal first-order rotation rule preserves a strict ordering of workers once the initial order is formed.

\citet{eliaz2026clerks} study assignment design under commitment in which workers privately observe task completion, may delay release of completed tasks, and cannot be replaced before release. Their baseline model concerns motivating release; they also study a two-worker hidden-effort extension. They exploit the no-replacement-before-release restriction to derive a payoff upper bound for the designer that their optimal design attains. In our model, this restriction is absent because the manager may reassign a worker after either public outcome of success or failure. Moreover, under hidden effort with at least three workers, replacing a current assignee is subject to the difficulty imposed by workers' interdependent effort incentives, as described above, which our analysis addresses.

\citet{board2011relational} and \citet{andrews2016allocation} study dynamic multi-agent relationships whose optimal assignment rules exploit observable, time-varying heterogeneity and monetary transfers, both of which are absent in our setting. Interdependent effort incentives in our analysis contrast with dynamic single-agent assignment where the principal decides whether or not to assign to one agent, but not whom to assign (\citealp{li2017power}; \citealp{bird2019dynamic}; \citealp{guo2020dynamic}; \citealp{lipnowski2020repeated}). Our analysis also contrasts with models of dynamic moral hazard in teams in which workers are complementary and there are no assignment dynamics (\citealp{che2001optimal}; \citealp{rayo2007relational}; \citealp{bonatti2011collaborating}; \citealp{georgiadis2015projects}).

Several adjacent papers further clarify our contribution. \citet{watt2026paying} studies dynamic assignment of tasks of heterogeneous difficulty and focuses on inducing workers to accept unpleasant assignments rather than on inducing hidden effort.  \citet{mylovanov2008task} study task assignment to ex ante identical agents in a two-period moral-hazard model with transfers. \citet{butler2017motivating} study favorable task assignments as status prizes in a dynamic tournament; we instead optimize over all assignment rules in our setting and our optimal rule does not rely on relative-performance comparisons.

Our analysis complements three strands of the literature on organizational economics. The first studies how assignment exploits heterogeneity in workers' skills, information, or productivity, rather than motivating hidden effort; the second studies rotation that is not contingent on performance, which provides no effort incentives in our model; see, \egg \citet{garicano2013hierarchies} for a survey. Finally, the discriminatory nature of first-order rotation broadens the effort-equality tension previously identified in static monetary contracts among complementary, ex ante identical workers (\citealp{winter2004incentives}; \citealp{halac2021rank}). Our setting is substantially different: the tension here is dynamic, self-enforcing, nonmonetary, and arises among technologically independent workers.

\section{Model}
\label{sec:model}

Time $t=0,1,\dots$ is discrete and the horizon is infinite. A manager interacts with a set of ex ante identical workers $N: = \{ 1, \dots, n \}$, with $n \ge 1$. The manager has a task that is undesirable for the workers.

In each period, the manager either leaves the task vacant or publicly assigns one worker to it. The assignee privately chooses whether to exert effort or shirk. This action yields a noisy public outcome: success $S>0$ or failure $F<0$. Effort yields success with probability $p\in(0,1)$ and failure otherwise; shirking yields success with probability $q\in(0,p)$ and failure otherwise. At the beginning of the period, players observe the realization of a public randomization device capable of implementing any prescribed finite lottery.

In each period $t$, each worker $i$'s payoff, $u^i_t$, and the manager's expected payoff, $v_t$, are as follows. The assignee receives a payoff normalized to zero if he exerts effort and a payoff gain $s>0$ if he shirks. Each unassigned worker receives a ``resting'' payoff $r > 0$. The manager receives a realized payoff of $S$ after success, $F$ after failure, and zero if the task is left vacant. Her expected payoff is therefore $\overline v:=pS+(1-p)F$ when the assignee exerts effort and $\underline v:=qS+(1-q)F$ when he shirks. We assume $(p,q,S,F)$ are such that $\overline v>0$ and $\underline v<0$, so the manager prefers effortful assignment to vacancy and vacancy to assignment with shirking.

We abstract from workers' participation decisions to isolate effort incentives; \cref{sec:extensions} introduces an outside option and derives the corresponding participation constraints. The binary-action restriction is likewise inessential: \cref{sec:richeractions} extends our results to a continuum of effort choices. By contrast, restricting outcomes to be binary is important for our explicit equilibrium characterization. This assumption is standard in the principal-multi-agent literature and captures settings in which performance is assessed by the success or failure of an operational objective, as described in the Introduction.\footnote{\cref{app:ternary-rank-boundary} illustrates that with richer outcomes, conditioning distinct continuations on different outcome realizations may strengthen effort incentives, so attaining the manager's optimum in general requires more nuanced equilibrium constructions.} Finally, although public randomization is often modeled as an independent uniform draw from $[0,1]$, we take finite lotteries as primitive, since they suffice for our results and avoid unnecessary measurability details.

At the beginning of period $t$, the public history is a sequence $h_t \in ((N \times \{S,F\}) \cup \{\varnothing\})^t$ recording, for each previous period, either the assignee's identity and outcome if a worker was assigned or the symbol $\varnothing$ if the task was left vacant, as well as all past and current realizations of the public randomization device that, as is customary, are suppressed from the notation. Worker $i$'s strategy is a collection $\gs^i \equiv (\gs^i_t)^\infty_{t=0}$, where $\gs^i_t (h_t) \in [0,1]$ is the probability that worker $i$ exerts effort when assigned at history $h_t$.\footnote{The stage game is extensive-form. Equivalently, the realized assignment determines a publicly observed within-period state in which only the assignee has a nontrivial action. Our notation incorporates this state by specifying each worker's action conditional on being assigned in his strategy.} Thus, workers do not condition their strategies on past private actions. This restriction is innocuous for our results: \cite{fudenberg1994efficiency} show that in repeated games with public product-structure monitoring, such as ours, the set of sequential equilibrium payoffs coincides with that of perfect public equilibrium payoffs, although sequential-equilibrium strategies may depend on past private actions. The manager's strategy is a collection $f \equiv (f_t)^\infty_{t=0}$, where $f_t(h_t) \in \Delta(N \cup \{\varnothing\})$ specifies a distribution from which the current assignee, if any, is drawn at history $h_t$. Therefore, her assignment may depend on all past assignments and outcomes in arbitrary ways.

The workers and the manager share a common discount factor $\gd \in (0,1)$. Each worker $i$'s and the manager's (discounted-)average payoffs are
\begin{align*}
(1-\gd)\sum_{t=0}^\infty \gd^t u_t^i \quad \text{ and } \quad (1-\gd) \sum_{t=0}^\infty \gd ^t v_t.
\end{align*}

We use perfect public equilibrium \citep{fudenberg1994folk}, henceforth equilibrium, as the solution concept. An equilibrium exists: permanent vacancy, with workers shirking following any unexpected assignment, is an equilibrium. We focus on no-shirking equilibria in which each assignee exerts effort on path, and characterize those that maximize the manager's payoff. This restriction reflects our high-stakes applications, as motivated at the outset, and is without loss of generality when the manager's payoff from an assignment with shirking is sufficiently low or when her first-best payoff is attainable.\footnote{More precisely, one sufficient condition under which a no-shirking equilibrium attains the manager's maximum payoff over all equilibria is $\underline v \le -\gd\overline v/(1-\gd)$; see \cref{sec:oa-suff} for a proof. Another is $c\leq\overline c_n$, where both $c$ and $\overline c_n$ are defined in \eqref{eq:secondbest-definitions} below. By \cref{thm:secondbest}, this condition permits a no-vacancy, no-shirking equilibrium that attains the manager's first-best payoff and hence her maximum payoff over all equilibria. Outside these cases, we do not characterize the unrestricted optimum. Doing so requires a different analysis that lies beyond the scope of this paper. The standard approach would be recursive, treating workers' promised utilities as state variables and jointly characterizing equilibrium behavior and the manager's payoff frontier as a function of these promises (see, \egg \citealp{li2017power}). To our knowledge, existing analyses taking this approach focus primarily on single-agent problems, for which the state variable is one-dimensional. With multiple workers, by contrast, the state is multi-dimensional: the frontier depends jointly on all workers' continuation payoffs and must incorporate the interdependence of their incentives.} This restriction is also standard in the reputation literature (\egg \citealp{mailath2001}) and the contracting literature (\egg \citealp{demarzo2006optimal}). Throughout the paper, optimality and the manager's optimum refer to optimization over no-shirking equilibria. We say that a no-shirking equilibrium is nontrivial if some assignment occurs on path.

We call the manager's strategy an assignment rule. Since workers' actions are hidden, only the manager may have an observable deviation in equilibrium. The manager's on-path behavior can be sustained by an off-path specification in which workers shirk when assigned and the manager assigns no one, because such a specification gives the manager her minmax payoff. Hereafter, we omit mentioning off-path behavior for conciseness and focus on workers' incentives.

We do not impose a parameter restriction that effortful assignment maximizes total surplus relative to either shirking or vacancy in each period. If such a restriction were imposed, the manager's payoff and social surplus would both be (weakly) increasing in the frequency of effortful assignment within the class of no-shirking equilibria. Our main result would then characterize both a manager-optimal and a welfare-maximizing no-shirking equilibrium.

We state a preliminary result that assignment dynamics in any no-shirking equilibrium are governed by effort incentives that are intrinsically dynamic. In any equilibrium, denote worker $i$'s continuation payoff at history $h_t$ by $U^i(h_t) := \bE \!\left[ 
(1-\gd)\sum_{\tau=t}^\infty \gd^{\tau-t} u^i_\tau
\middle| h_t \right]$, where the expectation is taken with respect to the equilibrium outcome distribution. Let $h_t i$ be the history that is a concatenation of history $h_t$ followed by assignment of worker $i$, and let $h_t i y$ be the history that is a concatenation of history $h_ti$ followed by outcome $y$.

\begin{lem*}[Incentives]\label{lem:ic}
In any no-shirking equilibrium, at any history $h_t$ on path where worker $i$ is assigned with positive probability,
\begin{align}\label{eq:ic}
\gd (p-q) (U^i(h_t i S) - U^i(h_t i F)) &\ge  (1-\gd) s.
\end{align}
\end{lem*}

\cref{lem:ic} shows that each on-path assignee exerts effort only if success yields a sufficiently higher continuation payoff than failure. Since a worker's continuation payoff reflects his average frequency of being assigned in any no-shirking equilibrium, this condition requires each current assignee to face a sufficiently lower future workload after success than after failure.

\cref{sec:proofs} collects the proofs of all formal results in the main text.

\section{Optimal assignment}
\label{sec:rotation}

In this section, we present our main result, beginning with a definition and notation that are essential.

\begin{def*}[First-order rotation]
\label{def:rotation}
An assignment rule is a \emph{first-order rotation rule} if it operates in rotation phases separated by vacancy phases, beginning with a rotation phase. Throughout each rotation phase, workers have distinct labels $\pi_1,\dots,\pi_n$, and the worker labeled $\pi_n$ is assigned in each period. After failure, all labels remain unchanged. After success, the assignee is relabeled $\pi_1$, and each worker labeled $\pi_k$, for $k<n$, is relabeled $\pi_{k+1}$. The rotation phase ends after $n$ successes. At the beginning of the next period, the public randomization device selects a length $L\in\mathbb N\cup\{\infty\}$ of the ensuing vacancy phase according to a (finite) lottery whose distribution may depend on the public history preceding that draw. During this phase, no worker is assigned and all labels remain unchanged. Once the vacancy phase ends, a new rotation phase begins.
\end{def*}

Therefore, a first-order rotation rule retains each assignee after failure and relieves him after success until every other worker has produced a success and a vacancy phase whose length may be randomized has elapsed. It is first-order in the sense that each current assignee's retention or relief depends only on his current outcome. A higher index $k$ in $\pi_k$ denotes a lower rank, so the lowest-ranked worker is assigned in each period during a rotation phase. 

We adopt the convention $0 \in \bN$, and call a first-order rotation rule with $L=0$ after every rotation phase an uninterrupted first-order rotation rule.  The initial ordering of the workers is arbitrary and may be formed progressively at random during the first rotation phase: after each success, the next assignee may be selected according to an arbitrary distribution over workers who have not yet been selected in that phase; once every worker has been selected, the realized ordering is preserved thereafter.\footnote{Note that because the length of each vacancy phase, when it is randomly determined, is drawn according to the public randomization device, the manager's deviation from it remains observable.} 

Define
\begin{align}
c
:=
\frac{(1-\gd)s}{\gd(p-q)r},
\qquad
\beta
:=
\frac{\gd p}{1-\gd(1-p)},
\qquad
\overline c_n
:=
(1-\beta)
\frac{1-\beta^{n-1}}{1-\beta^n}.
\label{eq:secondbest-definitions}
\end{align}
By \cref{lem:ic}, $c$ is the minimum difference between the assignee's continuation payoffs after success and failure, normalized by $r$, required to induce effort in any equilibrium. It measures the severity of the moral hazard problem. By assumption, $c>0$. Next, $\gb$ is the expected discount factor under any no-shirking equilibrium with first-order rotation from a worker's first assigned period in a rotation phase to the period immediately following his success. Indeed, letting $T$ denote the (random) number of assigned periods up to and including a worker's success,  $\bE[\gd^T]=\sum_{t=1}^{\infty}p(1-p)^{t-1}\gd^t=\gb$. We call these assigned periods the worker's assignment spell. Finally, as \cref{thm:secondbest} below shows,  $\overline c_n$ is the largest value of $c$ for which uninterrupted first-order rotation can sustain a no-shirking equilibrium when there are $n$ workers.

\begin{thm*}[The optimum]
\label{thm:secondbest}
If $c \le 1 - \gb$, the manager's optimal payoff is 
\begin{align*}
\begin{cases}
~\overline v, & \quad \text{ if } c\le \overline c_n,\\[0.5em]
~\overline v \dfrac{(1-\beta^n)[(1-\beta)-\beta c]}{(1-\beta)^2},
& \quad \text{ if } \overline  c_n < c \le 1-\beta.
\end{cases}
\end{align*}
This payoff is attained by any no-shirking equilibrium with a first-order rotation rule in which the vacancy-phase length $L$ at the end of each rotation phase is selected as follows:
\begin{enumerate}\itemsep0em
\item[1.]
For $c\leq\overline c_n$, $L=0$.
\item[2.]
For $\overline c_n<c<1-\beta$, 
\begin{align*}
L
=
\begin{cases}
\ell,
&
\textnormal{with probability } \dfrac{
((1-\beta)-c)/(\beta^{n-1}[(1-\beta)-\beta c])
-\gd^{\ell+1}
}{
\gd^\ell-\gd^{\ell+1}
},
\\[0.5em]
\ell+1,
&
\textnormal{otherwise},
\end{cases}
\end{align*}
where $\ell$ is any nonnegative integer that satisfies
\begin{align}\label{eq:ellinequality}
\gd^{\ell+1}
\leq
\frac{
(1-\beta)-c
}{
\beta^{n-1}
\left[
(1-\beta)-\beta c
\right]
}
\leq
\gd^\ell.
\end{align}
\item[3.] For $c = 1-\gb$, $L=\infty$.
\end{enumerate}
Finally, if $c>1-\gb$, the manager's optimal payoff is zero,  attained by persistent vacancy.
\end{thm*}

\Cref{thm:secondbest} shows that among no-shirking equilibria, no alternative rule---despite the manager's flexibility to condition assignments on the public history---can deliver the manager a higher payoff. A single assignment structure attains this optimum over the parameter region for which a nontrivial no-shirking equilibrium exists, parameterized by the moral-hazard severity $c$---from uninterrupted operation to complete shutdown. If $c\leq\overline c_n$, then uninterrupted first-order rotation attains the first-best. If $\overline c_n<c<1-\beta$, the optimal first-order rotation rule appends a vacancy phase with possibly randomized length to each rotation phase, with vacancies being lengthier as the moral hazard problem becomes more severe. This randomization is independent and identical across phases; it allows the manager to provide exactly the expected discounted respite needed to motivate effort without imposing unnecessary vacancy at the expense of her payoff. In the lottery, the prescribed integer $\ell$ always exists: the expression between $\gd^{\ell+1}$ and $\gd^\ell$ in \eqref{eq:ellinequality} lies strictly between zero and one, while the sequence $(\gd^k)_{k =0}^\infty$ strictly decreases from one to zero. At $c=1-\beta$, the manager completes one rotation phase and then shuts down permanently. Finally, if $c>1-\beta$, the moral hazard problem is so severe that no effortful assignment can be sustained. 

Note that $\overline c_1=0$, whereas $\overline c_n>0$ for every $n\geq2$. Thus, sustaining the manager's first-best requires at least two workers. With only one worker, any first-best equilibrium, if one exists, assigns the task to him in every period, making his continuation payoff independent of the current outcome, violating the incentive constraint \eqref{eq:ic}. 

Intuitively, \cref{thm:secondbest} holds because the prescribed first-order rotation rule maximizes the aggregate average assignment frequency subject to the workers' interdependent effort incentives. This interdependence is particularly transparent in any equilibrium attaining the manager's first-best: the task is assigned in every period, so the workers' continuation payoffs sum to $(n-1)r$ at every on-path history. Increasing one worker's continuation payoff by reducing his workload therefore necessarily reduces the continuation payoff of at least one other worker, affecting the incentives available to motivate his effort.

Before sketching the proof, we clarify the scope of the theorem. The theorem implies that a nontrivial no-shirking equilibrium exists if and only if $0< c \le 1-\gb$. \cref{prop:rotation-necessity} below shows that within this region of parameters, first-order rotation is necessary on path for optimality only at the two cutoff values $c=\overline c_n$ and $c=1-\gb$. We say that two assignment rules coincide on path if they prescribe the same assignment or vacancy after every on-path history.

\begin{prop*}[Uniqueness and multiplicity]
\label{prop:rotation-necessity}
For $c \in (0,1-\gb]$, the following hold.
\begin{enumerate}\itemsep0em
\item If $c=\overline c_n$, every optimal no-shirking equilibrium coincides on path with uninterrupted first-order rotation.

\item If $c=1-\gb$, every optimal no-shirking equilibrium consists on path of one first-order rotation phase followed by permanent vacancy.

\item If $c\in (0,\overline c_n) \cup (\overline c_n,1-\gb)$, there exists an optimal no-shirking equilibrium whose on-path assignment rule is not first-order rotation.
\end{enumerate}
\end{prop*}

Intuitively, because uninterrupted first-order rotation sustains the manager's optimum for all $c\leq\overline c_n$, it provides stronger effort incentives than necessary when this inequality is strict. This slack permits other rules to attain the manager's first-best. For example, the manager can perturb first-order rotation by retaining the assignee after success with a sufficiently small probability. This weakens effort incentives without violating them, so uninterrupted effortful assignment remains sustainable. By contrast, if $\overline c_n<c<1-\gb$, the manager's first-best is unattainable, and every optimal no-shirking equilibrium must include vacancies on path. These vacancies create implementation slack: because incentives and the manager's payoff depend on average assignment frequencies, public randomization can reproduce the same frequencies through different realized assignment paths. No such slack exists at $c=\overline c_n$ and $c=1-\gb$. In the former case, the workers' effort incentive constraints bind under uninterrupted first-order rotation. In the latter, each assignee must receive the maximal possible reward following success---zero future workload---and the maximal feasible punishment following failure---immediate retention.

\section{Recursive upper bound}
\label{sec:proofsketch}

A central component of our proof of \cref{thm:secondbest} is a recursive upper bound. For each group size, we bound the largest aggregate average assignment frequency attainable by any group of workers of that size in any no-shirking equilibrium. Recursively iterating across group sizes yields an upper bound on the aggregate average assignment frequency of all workers and in turn the manager's payoff. In this section, we sketch this argument and explain how it leads to \cref{thm:secondbest}.

Fix any nontrivial no-shirking equilibrium. For each worker $i$ and any on-path history $h$, define 
\begin{align*}
x_i(h):=(1-\gd)\bE\!\left[\sum_{t=0}^{\infty}\gd^t\mathbf 1\{i\textnormal{ is assigned }t\textnormal{ periods after }h\}\middle|h\right]
\end{align*}
as his average assignment frequency in the continuation. Worker $i$'s continuation payoff is then $r (1-x_i(h))$. Let $A(h):=\sum_{i=1}^n x_i(h)$ denote the aggregate average assignment frequency in the continuation. The manager's continuation payoff is therefore $\overline v A(h)$. If worker $i$ is assigned at history $h$, his promise-keeping and incentive constraints can be expressed in terms of average assignment frequencies as follows:
\begin{align}
x_i(hi)
&=
(1-\gd)
+
\gd\left(
p x_i(hiS)+(1-p)x_i(hiF)
\right),
\label{eq:sb-promise-keeping}\\
x_i(hiF)-x_i(hiS)
&\geq c.
\label{eq:sbic}
\end{align}
We denote a generic on-path post-decision history as $h^*$, namely a history following the current assignment or vacancy decision. Let $B:=\max_{i\in N}\sup_{h^*}x_i(h^*)$ denote the supremum of individual average assignment frequencies over all workers and on-path post-decision histories. We first bound this supremum:

\begin{lem*}[Individual assignment-frequency bounds]\label{lem:individual-assignment-frequency}
It holds that
\begin{align}
c
\leq B
\leq
1-\frac{\gb}{1-\gb}c.
\label{eq:sb-individual-bound}
\end{align}
\end{lem*}

Because assignment occurs at some on-path history, $B>0$. To derive \eqref{eq:sb-individual-bound}, we choose a sequence of post-decision histories along which a fixed worker $i$'s average assignment frequency approaches $B$. If he is not currently assigned, his average assignment frequency in the continuation is at most $\gd B<B$, so he must eventually be assigned along this sequence. Write each such post-decision history as $hi$. Then, by definition of $B$, and because pre-decision average assignment frequencies are finite convex combinations of post-decision average assignment frequencies and obey the same bounds, the pre-decision history $hiF$ satisfies $x_i(hiF) \le  B$. Together with \eqref{eq:sbic}, it follows that
\begin{align}\label{eq:successfreq}
x_i(hiS) \le B-c.
\end{align}
Substituting this into \eqref{eq:sb-promise-keeping} and passing to the limit of the sequence of post-decision histories give the upper bound in \eqref{eq:sb-individual-bound}. On the other hand, nonnegativity of average assignment frequency and \eqref{eq:sbic} together imply $B\ge c$, \ie the lower bound in \eqref{eq:sb-individual-bound}. 

Combining these upper and lower bounds shows that a nontrivial no-shirking equilibrium exists only if $c \le 1-\gb$. For $c>1-\gb$, the manager's highest payoff among no-shirking equilibria is zero, attained by permanent vacancy.

In the rest of this section, assume $c \le 1-\gb$. The next lemma states our recursive bound: it bounds, for each group size $k=1,\dots,n$, the largest aggregate average assignment frequency attainable by any group of that size, namely 
\begin{align*}
H_k:=\max_{\substack{C\subseteq N\\|C|=k}}\sup_{h^*}\sum_{i\in C}x_i(h^*).
\end{align*}
Set $H_0:=0$.

\begin{lem*}[Recursive upper bound]\label{lem:group-assignment-frequency}
Suppose that $c \le 1-\gb$. For every $k=1,\ldots,n$,
\begin{align}
H_k \le
1-\frac{\gb}{1-\gb}c+\gb H_{k-1}.
\label{eq:Hkrecur}
\end{align}
\end{lem*}

To derive \eqref{eq:Hkrecur}, we choose a sequence of post-decision histories along which a fixed group of $k$ workers' total average assignment frequency approaches $H_k>0$. The group must eventually contain the current assignee; otherwise, its average assignment frequency would be at most $\gd H_k<H_k$. Because there are finitely many workers, some worker $i$ is assigned infinitely often along the sequence. Accordingly, without loss of generality, we restrict attention to a subsequence of those histories, in which worker $i$ is always the current assignee. Following success, his average assignment frequency is at most $B-c$ by \eqref{eq:successfreq}, while the other $k-1$ workers' aggregate average assignment frequency is at most $H_{k-1}$. Following failure, the group's aggregate average assignment frequency is at most $H_k$. Summing the group's promise-keeping conditions and taking the limit of the subsequence gives $H_k\leq(1-\gd)+\gd[p(H_{k-1}+B-c)+(1-p)H_k]$. Rearranging this inequality, and using the upper bound on $B$ from \eqref{eq:sb-individual-bound}, we obtain \eqref{eq:Hkrecur}. 

We then show that this recursion leads to an upper bound on the manager's payoff over nontrivial no-shirking equilibria. Again, because pre-decision average assignment frequencies are finite convex combinations of the corresponding post-decision average assignment frequencies and obey the same bounds, the aggregate average assignment frequency $A(h)$ at any (pre-decision) history $h$ is bounded above by $H_n$. For $c \le 1-\gb$, iterating \eqref{eq:Hkrecur} from $H_0=0$ over $k$ yields $A(h)\le H_n\le (1-\gb^n)[(1-\gb)-\gb c]/(1-\gb)^2$. Because $A(h)$ is at most one by definition, it is bounded above by
\begin{align}
\min\left(
1,
\frac{
(1-\gb^n)
\left[
(1-\gb)-\gb c
\right]
}{
(1-\gb)^2
}
\right).
\label{eq:sb-upper-bound}
\end{align}
Thus, for $c \le 1-\gb$, the manager's ex ante payoff is bounded by $\overline v$ times \eqref{eq:sb-upper-bound}. By direct computation, \eqref{eq:sb-upper-bound} equals one if $c\le \overline c_n$ and is strictly less than one otherwise.

\cref{thm:secondbest} then follows by verifying that the first-order rotation rule specified in the theorem constitutes a nontrivial no-shirking equilibrium attaining the manager's payoff upper bound above.

\section{Implications}
\label{sec:implications}

\Cref{thm:secondbest} has two substantive implications for organizational design. In this section, we discuss them in order.

\subsection{The first-best domain and comparative statics}
\label{sec:first-order-comparative-statics}

Our explicit characterization shows how staffing $n$, working conditions affecting the relief-assignment payoff gap $r$, and monitoring $(p,q)$ affect the manager's optimal payoff. Of particular interest is how they affect the first-best region $c\leq\overline c_n$, which is relevant when organizations cannot afford vacancies. 

We specify some essential notation. Under any strategy profile where the manager plays uninterrupted first-order rotation and assignees exert effort on path (which need not be an equilibrium), each worker's continuation payoff at each on-path history depends only on his label corresponding to that history; we let $U^{\mathrm{FB}}(\pi_k)$ denote a worker's continuation payoff when his label is $\pi_k$ at the beginning of a period, and often write $U^{\mathrm{FB}}(\pi_k)$ as $U^{\mathrm{FB}}(\pi_k;n,p,r)$ to stress the dependence on parameters $(n,p,r)$. Define 
\begin{align}\label{eq:icslack}
IC^{\mathrm{FB}}(n,p,q,r)
&:=
\gd(p-q)
\left[
U^{\mathrm{FB}}(\pi_1;n,p,r)-U^{\mathrm{FB}}(\pi_n;n,p,r)
\right]
-(1-\gd)s
\end{align}
as each assignee's incentive slack: by \cref{lem:ic}, uninterrupted first-order rotation sustains a no-shirking equilibrium if \eqref{eq:icslack} is nonnegative.\footnote{We suppress the dependence on $s$ and $\gd$ for brevity.} The first-best region $c \le \overline c_n$ can be equivalently written as $IC^{\mathrm{FB}}(n,p,q,r)\ge 0$.

\begin{prop*}[Comparative statics]\label{prop:Scs}
The following hold.
\begin{enumerate}
\item Let $n \ge 2$. $IC^{\mathrm{FB}}(n,p,q,r)$ is strictly increasing in $n$, $p$, and $r$, and strictly decreasing in $q$.\footnote{Although $\overline c_n-c$ and $IC^{\mathrm{FB}}$ have the same sign, their derivatives with respect to the parameters need not have the same sign: by direct computation, $IC^{\mathrm{FB}}(n,p,q,r)=\gd(p-q)r(\overline c_n-c)$, and the normalization factor $\gd r(p-q)$ depends on the parameters. We therefore study $IC^{\mathrm{FB}}$, which measures on-path assignees' effort incentives without this parameter-dependent normalization.}
\item It holds that
\begin{align}
\lim_{n\to\infty}\overline c_n
=
1-\gb.
\label{eq:workforce-cutoff-limit}
\end{align}
Consequently, because $c$ is independent of $n$, if $c<1-\gb$, a sufficiently large finite workforce allows the manager to attain her first-best.
\item The manager's optimal payoff is weakly increasing in $n$, $p$, and $r$, and weakly decreasing in $q$, holding the other primitives fixed. An increase in $p$ strictly raises her payoff if and only if a nontrivial no-shirking equilibrium exists at the perturbed parameter values. An increase in $n$ or $r$, or a decrease in $q$, strictly raises her payoff if and only if her first-best is unattainable at the original parameter values and a nontrivial no-shirking equilibrium exists at the perturbed parameter values.
\end{enumerate}
\end{prop*}

Part 1 shows that a larger workforce, a higher success probability under effort, a lower success probability under shirking, and more valuable relief all expand the first-best region. Under uninterrupted first-order rotation, a larger workforce lengthens respite following success, strengthening its value as a reward. The positive effect of a higher success probability under effort is nontrivial. Holding $q$ fixed, a higher $p$ shortens the assignee's own assignment spell and strengthens effort incentives directly, but it also accelerates rotation and shortens his respite following success, indirectly weakening effort incentives; the direct effect dominates. Finally, the intuition underlying the positive effect of more valuable relief and of a lower success probability under shirking is straightforward.

Part 2 strengthens the workforce comparative static: adding workers not only expands the first-best region but eventually allows the manager to attain her first-best whenever $c<1-\gb$---that is, whenever first-order rotation with finite vacancy phases can sustain a no-shirking equilibrium.

Part 3 extends these comparative statics to the manager's optimal payoff. A higher $p$ raises the manager's payoff from each effortful assignment and weakly increases the optimal assignment frequency. It therefore strictly raises her payoff whenever a nontrivial no-shirking equilibrium exists at the perturbed parameter values. By contrast, $n$, $q$, and $r$ affect her payoff only through the aggregate average assignment frequency, so they have a strict effect only when they strictly increase this frequency.

\subsection{The value of a persistent rotation order}
\label{sec:persistent-ordering}

\cref{thm:secondbest} identifies an incentive benefit of preserving a persistent rotation order. To highlight this benefit, consider a random-reselection benchmark inspired by \citet{de2021selecting}, as described in the Introduction, that replaces the queue in \cref{thm:secondbest} with random reselection, treating all workers except the most recent assignee symmetrically.

\begin{def*}[Random reselection]\label{def:random-reselection}
Let $n \ge 2$.\footnote{We exclude $n=1$. As will be clear from the definition, random reselection is not well-defined for $n=1$, in which case excluding the preceding rotation phase's last assignee leaves no worker eligible to begin the next one.} An assignment rule is a \emph{random-reselection rule} if it operates in rotation phases separated by vacancy phases, beginning with a rotation phase. The initial assignee is selected uniformly from all workers. Each assignee is retained until success. The rotation phase ends after $n$ successes. After a success that does not complete a rotation phase, the next assignee is selected uniformly from workers who have not yet succeeded in that phase. At the beginning of a vacancy phase, the public randomization device selects the vacancy-phase length $L\in\bN\cup\{\infty\}$ from a (finite) lottery whose distribution may depend on the public history preceding that draw. During this phase, no worker is assigned. Once the vacancy phase ends, a new rotation phase begins, with its first assignee selected uniformly from all workers except the preceding phase's last assignee.
\end{def*}

Therefore, under random reselection, the order of workers is selected anew in each phase, but immediate reselection of a worker who produced the most recent success is excluded. This rule coincides with a first-order rotation rule when $n=2$ but generally differs when $n>2$. Let $c^\dagger_n$ denote the largest value of $c$ for which some random-reselection rule sustains the manager's first-best.

\begin{prop*}[Value of persistent order]\label{prop:persistent-ordering}
Suppose $n>2$. Random reselection sustains a first-best equilibrium for the manager if and only if $c\leq c^\dagger_n$, where $0<c^\dagger_n<\overline c_n$; for such $c$, it attains the same payoff as her optimum in \cref{thm:secondbest}. Whenever $c^\dagger_n<c<1-\gb$, every no-shirking equilibrium under random reselection gives the manager a strictly lower payoff than her optimum in \cref{thm:secondbest}.
\end{prop*}

\Cref{prop:persistent-ordering} quantifies the advantage of first-order rotation over the random-reselection benchmark for the manager. Random reselection may return a successful worker to the task before every other worker has served since his success, shortening the respite generated by success. Preserving the queue eliminates this risk, strengthens incentives, and reduces the vacancy needed to sustain some effortful assignment.

\section{Extensions}
\label{sec:extensions}

In this section, we discuss several variants of our baseline model.

\paragraph{Perfect monitoring.} Our characterization of the manager's optimum in \cref{thm:secondbest} extends to perfect monitoring of workers' actions, $p=1$ and $q=0$. The upper bounds in \cref{lem:individual-assignment-frequency} and \cref{lem:group-assignment-frequency} continue to hold, using the worker's minmax payoff as a lower bound on his continuation payoff after an off-path failure.\footnote{By always shirking when assigned, a worker can guarantee himself $\min\{s,r\}$ at any history. An on-path assignee's payoff in a no-shirking equilibrium is at most $\gd r$, so a nontrivial no-shirking equilibrium requires $s\leq\gd r<r$. Consequently, $U^i(hiF)\geq s$, and the effort incentive constraint implies $U^i(hiS)\geq s/\gd$, or $x_i(hiS)\leq1-s/(\gd r)$. Promise keeping then gives $x_i(hi)\leq1-s/r$, and the group recursion yields $H_k\leq1-s/r+\gd H_{k-1}$.} The equilibrium construction also extends, with an explicit continuation following failure, which is now off path: retain the assignee until he exerts effort and succeeds, after which the prescribed rotation and vacancy phases continue.

\paragraph{Desirable task.} Our results extend to a desirable task, modeled by giving each unassigned worker payoff $-r<0$ rather than $r>0$. Assignment is then a reward rather than a burden, so first-order rotation reverses the outcome transitions: success retains the assignee, whereas failure relieves him, and each rotation phase ends after $n$ failures. Accordingly, the expected discount factor $\gb$ determining the recursion in \eqref{eq:Hkrecur} must be replaced by 
\begin{align}\label{eq:hatgb}
\widehat\gb
:=
\frac{\gd(1-p)}{1-\gd p},
\end{align}
since the retention probability is now $p$ rather than $1-p$ in the undesirable-task setting. The other arguments proceed analogously. See \cref{sec:desirable-task} for details.

\paragraph{Private payoff shocks.} Consider a variant of our baseline model in which workers face private idiosyncratic shocks to their cost of exerting effort. Suppose that in each period, each worker $i$ privately observes a shirking gain $s^i$ before choosing effort. These gains are independent across workers and time and are continuously distributed according to some possibly worker-specific distributions with identical support $[0,s]$. Our characterization extends with the baseline shirking gain $s$ interpreted as the upper endpoint. If the effort incentive constraint associated with $s$ fails, it also fails for realized gains sufficiently close to $s$. Conversely, satisfying that constraint makes effort optimal for every realized shirking gain.

\paragraph{Outside option.}
Consider a variant of our baseline model in which, in each period, the manager first publicly selects a worker, if any, for assignment. After observing this selection, all workers simultaneously decide whether to remain or exit permanently. If one or more workers exit, each exiting worker receives the normalized continuation payoff $\rho \in \bR$, while the remaining players receive a stage payoff of $0$ and the game proceeds to the next period.\footnote{The normalization of the stage payoff of an assignee who remains and exerts effort to zero remains innocuous, because the relevant payoff comparison for each worker's decision to stay or exit at each history is that between $\rho$ and his (expected) continuation payoff conditional on staying.} If no worker exits, the stage game proceeds as in the baseline model. The game ends once all workers have exited. Our results extend to no-exit no-shirking equilibria, namely no-shirking equilibria in which no worker exits on path, provided that the outside option $\rho$ is sufficiently low. Any such equilibrium induces the same on-path outcomes as the corresponding no-shirking equilibrium in the baseline model. Conversely, a baseline no-shirking equilibrium remains sustainable as a no-exit equilibrium if every worker's continuation payoff from remaining is at least $\rho$ after every on-path selection. For $c \le 1-\gb$, a worker's lowest continuation payoff under our first-order-rotation construction in \cref{thm:secondbest} is attained when he is selected for assignment. It therefore suffices that this payoff be at least $\rho$.\footnote{By direct computation, the requirement is
\begin{align*}
\rho \le
\begin{cases}
\dfrac{r\beta\overline c_n}{1-\beta},
& c\leq\overline c_n,\\[6pt]
\dfrac{r\beta c}{1-\beta},
& \overline c_n<c\leq1-\beta,\\[6pt]
r,
& c>1-\beta.
\end{cases}
\end{align*}
We do not characterize the manager's optimum subject to participation when this condition fails.
}

\paragraph{Limited worker information.}
In some applications, workers observe neither whether the task is assigned, nor the identity of the assignee, nor the resulting outcome, except when they themselves are assigned. Consider a variant of the baseline model with this information structure, given which strategies may depend on private histories; the model is otherwise unchanged. Adapting \citet{ely2005belief}, call an equilibrium belief-free if, after every private history of his own, each worker's continuation strategy is a best reply for every profile of the manager's and the other workers' histories consistent with his own.\footnote{\citet{ely2005belief} formulate belief-free equilibrium for two-player finite games, but their definition can be adopted here in the natural way.}

Uninterrupted first-order rotation constitutes a belief-free equilibrium that attains the manager's first-best payoff whenever $c \le \overline c_n$, as in the main model. By construction, conditional on being assigned, the success-failure continuation-payoff difference depends only on the number of other workers whose successful assignment spells must elapse before he is assigned again. It does not depend on his beliefs about the unobserved queue or the other workers' histories. The workers thus have no profitable deviation. Because the manager attains her first-best payoff in every continuation on path, she also has no profitable deviation. Similarly, when $c=1-\gb$, one round of first-order rotation followed by a shutdown constitutes a belief-free no-shirking equilibrium.

When $\overline c_n<c<1-\beta$, however, the optimal rule in \cref{thm:secondbest} prescribes finite vacancy phases. Here, the manager may be able to profit by deviating to shorten a vacancy phase without the workers detecting the deviation; the construction in the theorem therefore need not be sequentially rational for the manager. Commitment power can substitute for this lack of information: if the manager must commit to her strategy ex ante, then the first-order rotation with the vacancy lottery in \cref{thm:secondbest} can be implemented as a belief-free no-shirking outcome.

\section{Concluding comments}

This paper has characterized incentive provision through repeated task assignment among a group of ex ante identical workers in a pure moral hazard setting. We leave to future research the following extensions in which additional technological or informational constraints alter the assignment problem.

\paragraph{Multiple tasks per period.}
Suppose that multiple workers are assigned different tasks simultaneously in each period. Their outcomes jointly determine continuation play. Characterizing the manager’s first-best domain in this setting requires a multidimensional extension of our aggregate-assignment-frequency recursion.

\paragraph{Incomplete information.}
In applications where workers in the pool are less likely to be similarly qualified, they may possess private information about their ability. In the undesirable-task model, if a worker may be inherently incapable of producing success, sufficiently pessimistic beliefs about a worker's ability may lead the manager to relieve him after failure rather than retain him, weakening the punishment available to motivate effort. A systematic analysis of how the workers' private information shapes assignment dynamics requires a different analysis that tracks the evolution of the manager's and the workers' beliefs about each worker's ability.

\pagebreak
\appendix
\appendixtitleon

\begin{appendices}

\section{Proofs}
\label{sec:proofs}

\subsection{Proof of \cref{lem:ic}}

Fix a no-shirking equilibrium and on-path history $h_t$ where worker $i$ is assigned with positive probability. Worker $i$'s continuation payoff by exerting effort after being assigned is 
\begin{align*}
U^i(h_t i ) = (1-\gd) \times 0 + \gd ( p U^i(h_t i S )+ (1-p) U^i(h_t i F )).
\end{align*} By a one-shot deviation to shirk, this worker's continuation payoff is \begin{align*}
\hat U^i(h_t i ) = (1-\gd) \times s + \gd ( q U^i(h_t i S )+ (1-q) U^i(h_t i F )).\end{align*}
The one-shot deviation principle requires $U^i(h_t i ) - \hat U^i(h_t i )\ge 0$, 
which yields \eqref{eq:ic}.

\subsection{Proof of \cref{lem:individual-assignment-frequency}}

Fix a nontrivial no-shirking equilibrium and use the notation introduced in \cref{sec:proofsketch}. By assumption, $B>0$. Fix a worker $i$ attaining $B$ in the maximum over $N$. For every positive integer $m$, choose a post-decision history $h^*_m$ such that $x_i(h^*_m)>B-1/m$. If worker $i$ is not currently assigned along infinitely many elements of this sequence, then his assignment frequency along those elements is at most $\gd B$. Because $B-1/m>\gd B$ for all sufficiently large $m$, this is a contradiction. Hence, after discarding finitely many elements, worker $i$ is currently assigned; write the corresponding post-decision histories as $h_mi$. At each such history, \eqref{eq:sbic} and $x_i(h_miF)\leq B$ imply $x_i(h_miS)\leq B-c$. Substituting these inequalities into \eqref{eq:sb-promise-keeping} and taking the limit along the sequence gives $B\leq(1-\gd)+\gd[p(B-c)+(1-p)B]$. Rearranging and using $\gb/(1-\gb)=\gd p/(1-\gd)$ yields
\begin{align}
B
\leq
1-\frac{\gb}{1-\gb}c
=:
\overline B(c).
\label{eq:proof-oe-individual}
\end{align}
An effortful assignment also requires $B\geq c$. Indeed, \eqref{eq:sbic} and the nonnegativity of average assignment frequencies imply $x_i(hiF)\geq c$, whereas $x_i(hiF)\leq B$.

\subsection{Proof of \cref{lem:group-assignment-frequency}}

Fix an arbitrary no-shirking equilibrium with $c\leq1-\gb$ and use the notation introduced in \cref{sec:proofsketch}. For a given $k$, if $H_k=0$, the bound in \eqref{eq:proof-oe-recursion} below is immediate. Suppose that $H_k>0$, and fix a group $C$ attaining the maximum over groups of $k$ workers. For every positive integer $m$, choose a post-decision history $h^*_m$ such that $\sum_{j\in C}x_j(h^*_m)>H_k-1/m$. If the task is vacant or the current assignee does not belong to $C$ along infinitely many elements of this sequence, then the group's assignment frequency along those elements is at most $\gd H_k$. Because $H_k-1/m>\gd H_k$ for all sufficiently large $m$, this is a contradiction. Hence, after discarding finitely many elements, the current assignee belongs to $C$. Since $C$ is finite, pass to a subsequence along which the current assignee is the same worker $i\in C$ and write the corresponding histories as $h_mi$. Aggregating the promise-keeping constraints of the workers in $C$ at any such history gives
\begin{align*}
\sum_{j\in C}x_j(h_mi)
=
(1-\gd)+\gd\left[p\sum_{j\in C}x_j(h_miS)+(1-p)\sum_{j\in C}x_j(h_miF)\right].
\end{align*}
Following failure, the collection's total average assignment frequency is at most $H_k$. Following success, \eqref{eq:sbic} and $x_i(h_miF)\leq B$ imply $x_i(h_miS)\leq B-c$, while the remaining $k-1$ workers have total assignment frequency at most $H_{k-1}$. Taking the limit along the subsequence therefore gives
\begin{align*}
H_k
\leq
(1-\gd)+\gd[p(H_{k-1}+B-c)+(1-p)H_k].
\end{align*}
Because $\gb=\gd p/[1-\gd(1-p)]$ and $1-\gb=(1-\gd)/[1-\gd(1-p)]$, rearranging and applying \eqref{eq:proof-oe-individual} gives
\begin{align*}
H_k
\leq
(1-\gb)+\gb(H_{k-1}+B-c)
\leq
\gb H_{k-1}+(1-\gb)+\gb(\overline B(c)-c).
\end{align*}
The definition of $\overline B(c)$ in \eqref{eq:proof-oe-individual} implies $(1-\gb)+\gb(\overline B(c)-c)=\overline B(c)$. Hence,
\begin{align}
H_k
\leq
\overline B(c)
+
\gb H_{k-1},
\label{eq:proof-oe-recursion}
\end{align}
where $\overline B(c)$ is given in \eqref{eq:proof-oe-individual}.

\subsection{Proof of \cref{thm:secondbest}}\label{sec:mainproof}

Fix an arbitrary no-shirking equilibrium and use the notation introduced in \cref{sec:proofsketch}. The manager's continuation payoff is $\overline v A(h)$. Because at most one worker is assigned in each period,
\begin{align}
A(h)
\leq
1.
\label{eq:proof-oe-feasibility}
\end{align}
By \cref{lem:individual-assignment-frequency}, a nontrivial no-shirking equilibrium exists only if
\begin{align}
c
\le
1-\gb.
\label{eq:proof-oe-positive-assignment}
\end{align}
Therefore, if $c > 1-\gb$, the manager's payoff is zero, attained by permanent vacancy on path, with workers shirking following any unexpected assignment.

Suppose henceforth that $c\leq1-\gb$. Iterating \eqref{eq:proof-oe-recursion} in the proof of \cref{lem:group-assignment-frequency} from $H_0=0$ gives
\begin{align*}
H_n
\leq
\overline B(c)\sum_{j=0}^{n-1}\gb^j
=
\frac{\left(1-\frac{\gb c}{1-\gb}\right)(1-\gb^n)}{1-\gb}
=
\frac{(1-\gb^n) ((1-\gb)-\gb c)}{(1-\gb)^2}.
\end{align*}
Because $A(h^*)\leq H_n$ at every post-decision on-path history $h^*$, while pre-decision average assignment frequencies are finite convex combinations of post-decision average assignment frequencies, this bound and \eqref{eq:proof-oe-feasibility} imply that for every (pre-decision) history $h$ on path,
\begin{align}
A(h)
\leq
\min\left(
1,\,
\frac{
(1-\gb^n) ((1-\gb)-\gb c)
}{
(1-\gb)^2
}
\right).
\label{eq:proof-oe-upper}
\end{align}
The second term in \eqref{eq:proof-oe-upper} is at least one if and only if $(1-\gb^n)[(1-\gb)-\gb c]\geq(1-\gb)^2$, or equivalently, $c\leq(1-\gb)(1-\gb^{n-1})/(1-\gb^n)=\overline c_n$. Thus, effortful assignment on path in every period, and hence the manager's first-best, is possible only if $c\leq\overline c_n$.

Because the manager's continuation payoff is $\overline v A(h)$ at every on-path history $h$, to prove that the assignment rule stated in the theorem is optimal, it suffices to show that it attains the bound in \eqref{eq:proof-oe-upper}. Fix an arbitrary ordering of the workers, and prescribe effort at every assignment. Write $z:=\bE[\gd^L]$ and $\gd^\infty:=0$. Let $T$ denote the (random) number of periods in an assignment spell. As shown in \cref{sec:rotation}, $\bE[\gd^T]=\gb$. The average assignment frequency accumulated during the spell is therefore $(1-\gd)\bE[\sum_{t=0}^{T-1}\gd^t]=1-\bE[\gd^T]=1-\gb$.

Let $b(z)$ be a worker's assignment frequency at the beginning of his own assignment spell. From the beginning of one such spell to the beginning of his next spell, the rotation passes through $n$ successful assignment spells and one vacancy phase. Therefore, $b(z)=(1-\gb)+z\gb^n b(z)$, and hence $$b(z)=\frac{1-\gb}{1-z\gb^n}.$$ Following failure, the assignee remains assigned with the same continuation rule, so his continuation assignment frequency is $b(z)$. Following success, he waits through the other $n-1$ workers' assignment spells and the vacancy period before being assigned again. His continuation assignment frequency is therefore $z\gb^{n-1}b(z)$. The failure-success difference in assignment frequencies is
\begin{align}
D(z)
:=
b(z)-z\gb^{n-1}b(z)
=
(1-\gb)
\frac{1-z\gb^{n-1}}{1-z\gb^n}.
\label{eq:proof-oe-spread}
\end{align}
Thus, by \eqref{eq:sbic}, every assignee exerts effort whenever $D(z)\geq c$.

At the beginning of a rotation phase, the workers' assignment frequencies are $b(z),\gb b(z),\dots$, and $\gb^{n-1}b(z)$. The aggregate assignment frequency is consequently
\begin{align}
\widehat A(z)
:=
b(z)
\sum_{j=0}^{n-1}\gb^j
=
\frac{1-\gb^n}{1-z\gb^n}.
\label{eq:proof-oe-total}
\end{align}

We next verify effort incentives on path. Suppose first that $c\leq\overline c_n$. Here, $L=0$, so that $z=1$. Equation \eqref{eq:proof-oe-spread} gives $D(1)=(1-\gb)(1-\gb^{n-1})/(1-\gb^n)=\overline c_n\geq c$. Thus, every assignee exerts effort. Moreover, \eqref{eq:proof-oe-total} gives $\widehat A(1)=1$. Uninterrupted first-order rotation therefore attains the first-best.

Now suppose that $\overline c_n<c<1-\gb$. We verify that the vacancy lottery in the construction in \cref{thm:secondbest} is chosen such that each on-path assignee's effort incentive constraint binds, namely $D(z)=c$. Solving this equation using \eqref{eq:proof-oe-spread} gives $z(c):=\frac{(1-\gb)-c}{\gb^{n-1}[(1-\gb)-\gb c]}$. Both the numerator and denominator are positive because $c<1-\gb$. Moreover, $z(c)<1$ is equivalent to $c>(1-\gb)(1-\gb^{n-1})/(1-\gb^n)=\overline c_n$. Thus, $z(c)\in(0,1)$.

To implement this value, choose $\ell\in\bN$ such that $\gd^{\ell+1}\leq z(c)\leq\gd^\ell$ and set $\alpha:=[z(c)-\gd^{\ell+1}]/[\gd^\ell-\gd^{\ell+1}]\in[0,1]$. The lottery in \cref{thm:secondbest} selects $L=\ell$ with probability $\alpha$ and $L=\ell+1$ otherwise. It therefore satisfies $\bE[\gd^L]=\alpha\gd^\ell+(1-\alpha)\gd^{\ell+1}=z(c)$. By construction, $D(z(c))=c$, so every assignee's effort incentive constraint binds.

It remains to verify that this rule attains the manager's payoff upper bound. The expression for $z(c)$ gives $1-z(c)\gb^n=(1-\gb)^2/[(1-\gb)-\gb c]$. Substituting into \eqref{eq:proof-oe-total} yields $\widehat A(z(c))=(1-\gb^n)[(1-\gb)-\gb c]/(1-\gb)^2$, which equals the second term in \eqref{eq:proof-oe-upper}.

If $c=1-\gb$, then $L=\infty$, so that $z=0$. The manager completes one rotation phase and then leaves the task vacant forever. Equations \eqref{eq:proof-oe-spread} and \eqref{eq:proof-oe-total} give $D(0)=1-\gb=c$ and $\widehat A(0)=1-\gb^n$. The latter equals the second term in \eqref{eq:proof-oe-upper} evaluated at $c=1-\gb$, so the rule again attains the upper bound.

\subsection{Proof of \cref{prop:rotation-necessity}}

We first state some useful preliminaries. Let $0<c\leq1-\gb$. Use the notation of \cref{sec:proofsketch}, and recall $\overline B(c)$ given in \eqref{eq:proof-oe-individual}. Define $K_k(c):=\overline B(c)\sum_{j=0}^{k-1}\gb^j$ for $k=1,\ldots,n$, and set $K_0(c):=0$. By \cref{lem:individual-assignment-frequency} and \cref{lem:group-assignment-frequency}, each worker's average assignment frequency is bounded above by $\overline B(c)$, and each group of $k$ workers has total average assignment frequency bounded above by $K_k(c)$. These bounds hold at every pre-decision and post-decision history on path.

Suppose that a nonempty group $C$ of $k$ workers attains $K_k(c)$ at an on-path post-decision history. The current assignee must belong to $C$; otherwise, the group's frequency would be at most $\gd K_k(c)<K_k(c)$. Denote the assignee by $i$ and the history by $hi$. Following failure, the group's total average assignment frequency is at most $K_k(c)$. Following success, the assignee's frequency is at most $\overline B(c)-c$, while the remaining workers' total frequency is at most $K_{k-1}(c)$. Substituting these three bounds into the group's promise-keeping equation gives $K_k(c)\leq(1-\gd)+\gd[p(K_{k-1}(c)+\overline B(c)-c)+(1-p)K_k(c)]=K_k(c)$. Since both outcomes have positive probability, any strict inequality in these three bounds would make the group's current frequency strictly less than $K_k(c)$. All three bounds must therefore hold with equality. In particular, $x_i(hiS)=\overline B(c)-c$, so the effort incentive constraint gives $x_i(hiF)\geq\overline B(c)$. Together with $x_i(hiF)\leq\overline B(c)$, this yields
\begin{align}
x_i(hiF)&=\overline B(c), \qquad x_i(hiS)=\overline B(c)-c,
\label{eq:necessity-individual-equalities}\\
\sum_{j\in C}x_j(hiF)&=K_k(c), \qquad \sum_{j\in C\setminus\{i\}}x_j(hiS)=K_{k-1}(c).
\label{eq:necessity-set-equalities}
\end{align}
Because $x_i(hiF)$ averages post-decision frequencies that cannot exceed $\overline B(c)$, its equality to $\overline B(c)$ requires equality after every on-path realization of the next period's public lottery and assignment decision; the same reasoning applies to both group equalities in \eqref{eq:necessity-set-equalities}. Following failure, worker $i$ must therefore be retained, since otherwise his frequency would be at most $\gd\overline B(c)<\overline B(c)$. Following success, $C\setminus\{i\}$ attains its bound after every such realization. Inductively, the next $k$ assignment spells, whenever reached, have distinct assignees from $C$, each retained until success, with no intervening vacancies. 

We prove the three parts in order:

\begin{enumerate}\itemsep0em
\item Suppose that $c=\overline c_n$. Since $K_n(c)=1$, optimality requires uninterrupted assignment and aggregate average assignment frequency one at every on-path continuation. Applying the above preliminary observation with $C=N$ at the beginning of every assignment spell shows that any $n$ consecutive spells have distinct assignees. Write $w_m$ for the assignee in spell $m$. Whenever the relevant spells are reached, $\{w_m,\ldots,w_{m+n-1}\}=N$ and $w_{m+n}\notin\{w_{m+1},\ldots,w_{m+n-1}\}$, so $w_{m+n}=w_m$. Thus, the initial ordering may be determined progressively, but thereafter it persists without vacancies.

\item Suppose that $c=1-\gb$. Then $\overline B(c)=c$ and $K_k(c)=1-\gb^k$. Optimality requires initial aggregate average assignment frequency $K_n(c)$, so $C=N$ attains $K_n(c)$ after every initial on-path public realization and assignment decision. The above preliminary observation gives $n$ consecutive assignment spells with distinct assignees, each retained until success. Moreover, \eqref{eq:necessity-individual-equalities} gives every successful assignee zero continuation assignment frequency. After all $n$ successes, every worker has zero continuation assignment frequency, so the task remains vacant forever.

\item We construct alternative optimal equilibria in the two open regions $(0, \overline c_n)$ and $(\overline c_n, 1-\gb)$. Suppose first that $0<c<\overline c_n$, which requires $n\geq2$. Fix a cyclic ordering of workers and let $\alpha:=(1+c/\overline c_n)/2\in(0,1)$. Retain the assignee after failure; after success, advance to the next worker with probability $\alpha$ and retain the assignee otherwise. Never leave the task vacant. Then, provided that assignees exert effort on path, advancement occurs with probability $\alpha p$. Write $\gb_\alpha:=\gd\alpha p/[1-\gd(1-\alpha p)]$. The current assignee's average assignment frequency satisfies $b_\alpha=(1-\gb_\alpha)+\gb_\alpha^n b_\alpha$, giving $b_\alpha=(1-\gb_\alpha)/(1-\gb_\alpha^n)$. His continuation average assignment frequencies after failure and success are respectively $b_\alpha$ and $[(1-\alpha)+\alpha\gb_\alpha^{n-1}]b_\alpha$. Their difference is $D_n(\alpha)=\alpha(1-\gb_\alpha)(1-\gb_\alpha^{n-1})/(1-\gb_\alpha^n)$. Since $x\mapsto(1-x)(1-x^{n-1})/(1-x^n)$ is strictly decreasing on $(0,1)$ and $\gb_\alpha<\gb$, we have $D_n(\alpha)>\alpha\overline c_n=(\overline c_n+c)/2>c$. Thus effort is optimal (for each assignee on path) and uninterrupted assignment attains the first-best. The rule is not first-order rotation because success followed by retention has positive probability.

Now suppose that $\overline c_n<c<1-\gb$, allowing any $n\geq1$. We modify the optimal first-order rotation rule after its first completed rotation phase. The modification preserves every worker's expected continuation average assignment frequency but allows permanent vacancy following failure. We first construct a continuation with this property and then explain how to incorporate it into the optimal construction in \cref{thm:secondbest}.

Consider a continuation assigning only one designated worker. Retain him after failure with probability $c/d$, where $d:=(1-\gd)+\gd(1-p)c$, and leave the task vacant forever otherwise; after success, leave it vacant forever. Since $d-c=[1-\gd(1-p)](1-\gb-c)>0$, the retention probability lies in $(0,1)$. His average assignment frequency is the unique solution to $x=(1-\gd)+\gd(1-p)(c/d)x$, namely $x=d$. His continuation average assignment frequencies after failure and success are therefore $c$ and zero, respectively, so effort is optimal by \eqref{eq:sbic}.

We next identify the continuation frequencies that the modification must preserve. Let $z(c)\in(0,1)$ be as defined in \cref{sec:mainproof}, and label workers $w_0,\ldots,w_{n-1}$ in the order in which the original first-order rotation rule would assign them in the next rotation phase. Since $b(z(c))=\overline B(c)$, beginning the optimal continuation from \cref{thm:secondbest} immediately gives worker $w_j$ average assignment frequency $\gb^j\overline B(c)$. The original vacancy lottery therefore gives the expected continuation frequencies
\begin{align}
z(c)\overline B(c)(1,\gb,\ldots,\gb^{n-1}).
\label{eq:standard-postrotation-assignment-frequencies}
\end{align}
Choose $\varepsilon\in(0,\min\{z(c),(1-z(c))d/K_n(c)\})$. At the end of the first rotation phase, replace the vacancy lottery as follows: with probability $z(c)-\varepsilon$, begin the original optimal continuation immediately; with probability $\varepsilon\overline B(c)\gb^j/d$, implement the single-worker continuation for $w_j$, for each $j=0,\ldots,n-1$; with the remaining probability $1-z(c)+\varepsilon-\varepsilon K_n(c)/d$, leave the task vacant forever. Under this lottery, worker $w_j$'s expected continuation average assignment frequency is $(z(c)-\varepsilon)\overline B(c)\gb^j+(\varepsilon\overline B(c)\gb^j/d)d=z(c)\overline B(c)\gb^j$. Thus, the lottery preserves every worker's frequency in \eqref{eq:standard-postrotation-assignment-frequencies}. It therefore also preserves every worker's continuation payoff and the manager's continuation payoff at every history at which the first rotation phase ends.

Play before the end of the first phase is unchanged. Because the continuation payoffs at every possible completion history are also unchanged, each worker's continuation payoffs following success and failure at every earlier assignment history remain unchanged. Hence, all earlier effort incentive constraints continue to hold. Effort is optimal in every new continuation, and the manager's initial payoff is unchanged, so the resulting equilibrium is optimal. Its assignment rule is not first-order rotation: with positive probability, a single-worker continuation is reached and the task becomes permanently vacant following failure, whereas first-order rotation requires retention after failure.
\end{enumerate}

\subsection{Proof of \cref{prop:Scs}}\label{app:comparative-statics}

We prove the three parts in order.

\begin{enumerate}\itemsep0em
\item The payoffs $\{U^{\mathrm{FB}}(\pi_k): k=1,\dots,n\}$ satisfy
\begin{align*}
U^{\mathrm{FB}}(\pi_n)
&=
\gd\left[pU^{\mathrm{FB}}(\pi_1)+(1-p)U^{\mathrm{FB}}(\pi_n)\right],
\\
U^{\mathrm{FB}}(\pi_k)
&=
(1-\gd)r
+
\gd\left[pU^{\mathrm{FB}}(\pi_{k+1})+(1-p)U^{\mathrm{FB}}(\pi_k)\right],
\qquad k<n.
\end{align*}
By direct computation,
\begin{align}
IC^{\mathrm{FB}}(n,p,q,r)
&=
\gd r(p-q)\overline c_n-(1-\gd)s.
\label{eq:scope-expanded}
\end{align}

For fixed $\gb\in(0,1)$, \begin{align}
\overline c_{n+1}-\overline c_n
&=
\frac{
(1-\gb)^3\gb^{n-1}
}{
(1-\gb^n)(1-\gb^{n+1})
}
>0.
\label{eq:workforce-cutoff-difference}
\end{align} It follows from \eqref{eq:scope-expanded} that $IC^{\mathrm{FB}}$ is strictly increasing in $n$. Because $\gb$, and hence $\overline c_n$, is independent of $q$ and $r$, $\partial IC^{\mathrm{FB}}/\partial r=\gd(p-q)\overline c_n>0$ and $\partial IC^{\mathrm{FB}}/\partial q=-\gd r\overline c_n<0$.

It remains to establish strict monotonicity in $p$. Define $a:=(1-\gd)/\gd$ and $\rho_n(x):=(1-x^{n-1})/(1-x^n)$. Then $\gb=p/(p+a)$, and
\begin{align}
(p-q)\overline c_n
&=
\left[
a\gb-q(1-\gb)
\right]\rho_n(\gb)
\notag\\
&=
a\gb\rho_n(\gb)-q\overline c_n.
\label{eq:scope-p-representation}
\end{align}
The map $x\mapsto(1-x)\rho_n(x)=(1-x^{n-1})/(1+x+\cdots+x^{n-1})$ is strictly decreasing on $(0,1)$. Moreover, $\frac{\diff}{\diff x}[x\rho_n(x)]=[1-nx^{n-1}+(n-1)x^n]/(1-x^n)^2>0$. Indeed, the numerator is strictly decreasing on $(0,1)$ and equals zero at $x=1$. Thus both terms on the right-hand side of \eqref{eq:scope-p-representation} are strictly increasing in $\gb$. Because $\partial\gb/\partial p=a/(p+a)^2>0$, $(p-q)\overline c_n$, and hence $IC^{\mathrm{FB}}$, is strictly increasing in $p$.
\item By \eqref{eq:workforce-cutoff-difference}, $\overline c_n$ is strictly increasing in $n$. Moreover,
\begin{align*}
\lim_{n\to\infty}\overline c_n
&=
(1-\gb)
\lim_{n\to\infty}
\frac{1-\gb^{n-1}}{1-\gb^n}
=
1-\gb,
\end{align*}
which proves \eqref{eq:workforce-cutoff-limit}. Since $\overline c_n$ increases to $1-\gb$, any $c<1-\gb$ satisfies $c\leq\overline c_n$ for all sufficiently large finite $n$. By \cref{thm:secondbest}, such a workforce allows the manager to attain her first-best.
\item Let $A_n^*$ denote the manager's maximum average assignment frequency among no-shirking equilibria. Define $\kappa:=1-ps/[(p-q)r]$ and $R_n(\gb):=\sum_{j=0}^{n-1}\gb^j$. Since $\gb c/(1-\gb)=1-\kappa$, \cref{thm:secondbest} gives
\begin{align}
A_n^*
&=
\begin{cases}
0,
&
\kappa<1-\gb,
\\
\min\{1,\kappa R_n(\gb)\},
&
\kappa\geq1-\gb.
\end{cases}
\label{eq:optimal-frequency-kappa}
\end{align}
The manager's optimal payoff is $\overline v A_n^*$. Consider first an increase in $n$ or $r$, or a decrease in $q$, holding the other primitives fixed. These changes leave $\gb$ and $\overline v$ unchanged. Increasing $n$ strictly increases $R_n(\gb)$, while increasing $r$ or decreasing $q$ strictly increases $\kappa$. Thus, whenever $\kappa\geq1-\gb$ before the change, this inequality continues to hold afterward and $\kappa R_n(\gb)$ strictly increases. By \eqref{eq:optimal-frequency-kappa}, an original first-best payoff therefore remains unchanged, whereas an original positive payoff below the first-best strictly increases. If the original payoff is zero, it strictly increases if and only if a nontrivial no-shirking equilibrium exists at the perturbed parameter values. These cases establish both the weak comparisons and the stated necessary and sufficient conditions for strictness. Now consider an increase in $p$, holding the other primitives fixed. Both $\kappa$ and $\gb$ strictly increase, while $R_n(\gb)$ weakly increases. If a nontrivial no-shirking equilibrium exists at the original parameter values, then $\kappa\geq1-\gb>0$, this inequality continues to hold after the change, and \eqref{eq:optimal-frequency-kappa} implies that $A_n^*$ weakly increases. Moreover, $\overline v=pS+(1-p)F$ strictly increases because $S>F$, so the manager's payoff strictly increases. If no nontrivial no-shirking equilibrium exists at the original parameter values, her original payoff is zero; it becomes positive precisely when such an equilibrium exists at the perturbed parameter values. Thus, her payoff weakly increases in $p$, with a strict increase if and only if a nontrivial no-shirking equilibrium exists at the perturbed parameter values.
\end{enumerate}

\subsection{Proof of \cref{prop:persistent-ordering}}\label{sec:symm}
Suppose $n>2$. Consider a strategy profile where the manager plays uninterrupted random reselection and assignees exert effort. Let $w_k$ denote the continuation average assignment frequency following success of the worker selected $k$th in a rotation phase, for $k=1,\ldots,n$. Under this rule, $w_k$ depends only on $k$, $n$, $p$, and $\gd$. Because a successful worker is not assigned in the following period, $0\leq w_k\leq\gd<1$. After failure, the worker remains assigned until success. His continuation average assignment frequency is therefore $(1-\gb)+\gb w_k$, giving a failure-success difference of $(1-\gb)(1-w_k)$. By \eqref{eq:sbic}, effort is optimal at every assignment if and only if $c\leq(1-\gb)\min_{1\leq k\leq n}(1-w_k)$. Since any first-best equilibrium must have uninterrupted effortful assignment, it follows that $c^\dagger_n=(1-\gb)\min_{1\leq k\leq n}(1-w_k)>0$.

At $c=\overline c_n$, \cref{prop:rotation-necessity} requires every first-best equilibrium to follow uninterrupted first-order rotation on path. Thus, the worker who served first in the first rotation phase must also serve first in the second phase. Random reselection instead chooses the second phase's first assignee uniformly from the $n-1$ workers other than the preceding phase's last assignee. Only one of these workers served first in the preceding phase, so random reselection selects someone else with probability $(n-2)/(n-1)>0$. It therefore cannot sustain a first-best equilibrium at $c=\overline c_n$. Since we have shown that random reselection sustains the first-best if and only if $c\leq c^\dagger_n$, this implies $c^\dagger_n<\overline c_n$. For $c^\dagger_n<c\leq\overline c_n$, first-order rotation attains the manager's first-best, whereas no no-shirking equilibrium under random reselection does. Every such equilibrium therefore gives the manager strictly less than her optimum in \cref{thm:secondbest}.

It remains to establish the payoff comparison when $\overline c_n<c<1-\gb$. Fix any no-shirking equilibrium under random reselection. We first count only assignments within a single rotation phase and measure average assignment frequencies from the beginning of that phase. A worker selected in position $k$ waits for $k-1$ other workers to succeed before he is assigned. Each preceding assignment spell discounts his assignments by the expected discount factor $\gb$, while his own assignment spell has average frequency $(1-\gd)\bE[\sum_{t=0}^{T-1}\gd^t]=1-\bE[\gd^T]=1-\gb$, where $T$ is the random length of his spell. His average assignment frequency from assignments within this phase is therefore $(1-\gb)\gb^{k-1}$.

Consider any rotation phase after the first, before its first assignee is selected. The preceding phase's last assignee cannot be selected first and is equally likely to occupy positions $2,\ldots,n$. His average assignment frequency from assignments within this phase is therefore $(1-\gb)a_n$, where $a_n:=(n-1)^{-1}\sum_{j=1}^{n-1}\gb^j$. Each other worker may be selected first; conditional on not being selected first, he is also equally likely to occupy positions $2,\ldots,n$. Being selected first gives him average assignment frequency $1-\gb$, which exceeds $(1-\gb)a_n$. Thus, every worker's average assignment frequency from assignments within the phase is at least $(1-\gb)a_n$. The aggregate frequency is the sum across all positions, $(1-\gb)\sum_{j=0}^{n-1}\gb^j=1-\gb^n$.

Now fix a history $h$ immediately after a rotation phase ends and before the vacancy lottery. Measure time from the period immediately following $h$, and let $T_\ell$ be the beginning date of the $\ell$th subsequent rotation phase. Set $T_\ell=\infty$ if that phase never begins, with $\gd^\infty:=0$. Assignments in a rotation phase beginning at date $T_\ell$ receive the additional discount $\gd^{T_\ell}$ when measured from $h$ rather than from that phase's beginning.

The preceding calculation applies conditional on every history at the beginning of a phase, before its first assignee is selected. Taking expectations over these histories, assignments during phase $\ell$ therefore add $(1-\gb^n)\bE[\gd^{T_\ell}| h]$ to $A(h)$ and at least $(1-\gb)a_n\bE[\gd^{T_\ell}| h]$ to $x_i(h)$ for every worker $i$. Since every assignment after $h$ occurs in one of these phases, summing gives
\begin{align}
A(h)&=(1-\gb^n)\sum_{\ell=1}^{\infty}\bE[\gd^{T_\ell}| h],\notag\\
x_i(h)&\geq(1-\gb)a_n\sum_{\ell=1}^{\infty}\bE[\gd^{T_\ell}| h]
=\frac{(1-\gb)a_n}{1-\gb^n}A(h),\qquad i\in N.
\label{eq:symmetric-phase-bound}
\end{align}

Consider an assignment history $hi$ at which every worker other than the current assignee $i$ has already succeeded in the current phase. Success by $i$ therefore completes that phase. By \cref{lem:individual-assignment-frequency} and \eqref{eq:sbic}, $x_i(hiS)\leq x_i(hiF)-c\leq1-\gb c/(1-\gb)-c=(1-\gb-c)/(1-\gb)$. Applying \eqref{eq:symmetric-phase-bound} at $hiS$ yields $A(hiS)\leq(1-\gb^n)(1-\gb-c)/[(1-\gb)^2a_n]$. This bound holds after every possible rotation-phase-completion history.

Let $A_0$ denote the initial aggregate average assignment frequency. The first rotation phase contributes $1-\gb^n$, and the expected discount factor from its beginning to the period immediately following its completion is $\gb^n$. Hence,
\begin{align}
A_0
&\leq1-\gb^n+\frac{\gb^n(1-\gb^n)(1-\gb-c)}{(1-\gb)^2a_n}\notag\\
&<1-\gb^n+\frac{\gb(1-\gb^n)(1-\gb-c)}{(1-\gb)^2} =\frac{(1-\gb^n)[(1-\gb)-\gb c]}{(1-\gb)^2}.
\label{eq:symmetric-interior-gap}
\end{align}
The strict inequality follows from $a_n>\gb^{n-1}$ and $c<1-\gb$. The final expression in \eqref{eq:symmetric-interior-gap} is the optimal aggregate average assignment frequency in \cref{thm:secondbest}. Multiplying by $\overline v>0$ establishes the strict payoff comparison and completes the proof.

\pagebreak

\section{Omitted details}

\subsection{Richer actions}
\label{sec:richeractions}

Suppose instead that, when assigned, a worker chooses effort $a\in A\equiv[\underline a,\overline a]$. The manager's expected flow payoff is $v(a)$ when the assignee chooses $a$ and zero when the task is vacant. Assume that
\begin{align*}
\overline a\in\argmax_{a\in A}v(a),
\qquad
\overline v:=v(\overline a)>0,
\qquad
\underline v:=v(\underline a)<0.
\end{align*}
Call an equilibrium a target-effort equilibrium if every worker assigned on path chooses $\overline a$. We characterize the manager's optimum within this class.

Slightly abusing notation, let $p:A\ra(0,1)$, where $p(a):=\Pr(S| a)$ denotes the probability of success under effort $a$. The worker's stage-game payoff when assigned and choosing $a$ is $k(\overline a)-k(a)$, where $k$ is his effort-cost function. Assume that $p$ and $k$ are continuously differentiable on $A$, that $p$ is concave with $p'>0$, and that $k$ is increasing and convex. The additive constant $k(\overline a)$ ensures that the worker's payoff from choosing target effort $\overline a$ is again normalized to zero; lowering effort generates a private gain. Each unassigned worker continues to receive payoff $r>0$. With binary outcomes, $p'>0$ is the monotone likelihood-ratio property, while concavity of $p$ is the familiar convexity-of-the-distribution-function condition. Together with convexity of $k$, these conditions validate the first-order approach in this environment (see, \egg \citealp{rogerson1985first}).

Consider an on-path history $h_t$ at which worker $i$ is assigned. If worker $i$ chooses effort $a$, his continuation payoff from the current period onward is
\begin{align*}
(1-\gd)\left[k(\overline a)-k(a)\right]
+
\gd\left[
p(a)U^i(h_tiS)
+
(1-p(a))U^i(h_tiF)
\right].
\end{align*}
Dropping terms that are independent of $a$, the worker's payoff-relevant objective is
\begin{align*}
\Phi(a)
:=
\gd p(a)
\left[
U^i(h_tiS)-U^i(h_tiF)
\right]
-
(1-\gd)k(a).
\end{align*}
If $U^i(h_tiS)\geq U^i(h_tiF)$, then $\Phi$ is concave, so $\overline a$ maximizes $\Phi$ on $A$ if and only if the left derivative of $\Phi$ at $\overline a$ is nonnegative. If $U^i(h_tiS)<U^i(h_tiF)$, this left derivative is strictly negative, so $\overline a$ is not optimal. In either case, worker $i$'s incentive constraint for choosing $\overline a$ is
\begin{align}
U^i(h_tiS)-U^i(h_tiF)
\geq
\frac{(1-\gd)k'(\overline a)}
{\gd p'(\overline a)}.
\label{eq:rich-action-ic}
\end{align}

Therefore, the richer-action environment has exactly the same recursive assignment structure as the baseline model. The two changes are that $p$ is replaced by $p(\overline a)$ and that the discrete shirking-gain-to-informativeness ratio $s/(p-q)$ is replaced by the marginal-cost-to-local-informativeness ratio $k'(\overline a)/p'(\overline a)$.

\subsection{A three-signal illustration}
\label{app:ternary-rank-boundary}

This appendix provides the three-signal example discussed in the main text. Consider the model with an undesirable task and $n=3, r=1, \delta=\frac23$, and $s=\frac16$. The set of outcomes is $Z=\{z_L,z_M,z_H\}$, with distributions 
\begin{align*}
p_z:=\Pr(z|\textnormal{effort}) = \begin{cases}
~\frac 1 4, \quad &\text{ if }  z = z_L,\\
~\frac 1 2, \quad &\text{ if }  z = z_M,\\
~\frac 1 4, \quad &\text{ if }  z = z_H,
\end{cases}\quad 
q_z:=\Pr(z|\textnormal{shirk}) = \begin{cases}
~\frac 1 2, \quad &\text{ if }  z = z_L,\\
~\frac 3 8, \quad &\text{ if }  z = z_M,\\
~\frac 1 8, \quad &\text{ if }  z = z_H.
\end{cases}
\end{align*} 
Here, worker $i$'s incentive constraint for effort at history $h$ when assigned in any equilibrium becomes
\begin{align}\label{eq:genic}
\delta\sum_{z\in Z}(p_z-q_z)U^i(hiz) \ge (1-\delta)s.
\end{align}
In each period, continue to let the manager's expected payoffs from effort and shirking be $\overline v>0$ and $\underline v<0$, and let her payoff from a vacant period be $0$.

Consider an assignment rule that uses a ranking $(\pi_1,\pi_2,\pi_3)$ and assigns the bottom-ranked worker $\pi_3$ in each period on path. Following $z_H$, the rule reinserts the assignee at rank 1; following $z_M$, it reinserts him at rank 2; and following $z_L$, it leaves him at rank 3.  In every case, the rule preserves the relative order of the other two workers and repeats the same rule from the resulting ranking. In any no-shirking equilibrium with this assignment rule, each worker's continuation payoff at every history on path is a function of his label; let $U(\pi_k)$ be a worker's continuation payoff at rank $k$. It follows that the continuation payoffs are $(U(\pi_1), U(\pi_2), U(\pi_3))=(\frac 8 9, \frac 2 3, \frac 4 9)$ in any no-shirking equilibrium corresponding to this rule. The incentive constraint \eqref{eq:genic} holds, and both sides are equal to $\frac{1}{18}$. Thus the assignee is willing to exert effort at every ranking, and the graded rule supports a no-shirking equilibrium.

It remains to compare this rule with binary outcomes. To do so, we partition the current outcome space into two cells, with one cell representing success and the other representing failure as in the main text. For this comparison, continuation play depends on each outcome only through its binary report. For each partition, we label as success the cell that is more likely under effort than under shirking and write $p$ and $q$ as the probabilities of success under effort and shirking. There are three nontrivial binary partitions, given in the table below, subject to the restriction $p>q$ as in the baseline model:
\begin{align*}
\begin{array}{c|c|c|c}
\text{success} & p & q & I(p,q)\\ \hline
z_H       & 1/4 & 1/8 & 2/39\\[1mm]
z_M       & 1/2 & 3/8 & 1/28\\[1mm]
\{z_M,z_H\} & 3/4 & 1/2 & 8/147
\end{array}
\end{align*}
For $n=3$, $r=1$, and $\delta=2/3$, the first-order-rotation calculation in \cref{thm:secondbest} implies that a (manager's) first-best equilibrium under a given binary partition exists if and only if
\begin{align}\label{eq:specialic2}
I(p,q):=\frac23(p-q)\frac{1+4p}{1+6p+12p^2} \ge \frac{1}{18}.
\end{align}
The largest of the values $I(p,q)$ in the table is $\frac{8}{147} < \frac{1}{18}$, violating \eqref{eq:specialic2}. Consequently, no deterministic binary coarsening followed by first-order rotation supports the manager's first-best for these parameters, whereas the graded three-signal rule does.

The conclusion is unchanged if the coarsening is stochastic, assigning a probability of reporting success to each original outcome. For a given effort-side success probability $p$, likelihood-ratio ordering maximizes the difference between the effort- and shirking-side success probabilities by assigning success first to $z_H$, then to $z_M$, and finally to $z_L$. Thus, as $p$ increases, the optimal coarsening first randomizes after $z_H$ for $p\in[0, \frac 1 4]$, then after $z_M$ for $p\in[\frac 1 4, \frac 3 4]$, and finally after $z_L$ for $p\in[\frac 3 4,1]$. Substitution into $I$ shows that it is increasing over the first two ranges and decreasing over the third. Its maximum therefore occurs at $p=\frac 3 4$, corresponding to the deterministic coarsening $\{z_M, z_H\}$, and equals $\frac{8}{147}$.

\subsection{Optimality of no-shirking equilibrium}
\label{sec:oa-suff}

In this Appendix, we provide sufficient conditions under which a no-shirking equilibrium attains the manager's maximum payoff over all equilibria.

\begin{prop*}[Unrestricted optimum]\label{prop:unrestricted-optimum}
If $\underline v\leq-\gd\overline v/(1-\gd)$, then the manager's maximum payoff over all equilibria is attained by the no-shirking equilibrium characterized in \cref{thm:secondbest}.
\end{prop*}

\begin{proof}[Proof of \cref{prop:unrestricted-optimum}]
Let $\underline v\leq-\gd\overline v/(1-\gd)$. If $c\leq\overline c_n$, then \cref{thm:secondbest} provides a no-shirking equilibrium attaining the manager's first-best payoff $\overline v$, which is also an upper bound on her payoff over all equilibria. Suppose that $c>\overline c_n$.

Fix an arbitrary equilibrium. We adopt the notation for pre-decision and post-decision histories $h$ and $h^*$  in \cref{sec:proofsketch}. Let $V(h)$ denote the manager's continuation payoff. At every history preceding her assignment decision, the manager can guarantee herself zero by leaving the task vacant forever. Moreover, $V(h)\leq\overline v$ at every continuation history because her expected stage payoff never exceeds $\overline v$.

We first show that every on-path assignee exerts effort with positive probability. Suppose instead that worker $i$ is assigned with positive probability at some on-path history $h$ and shirks with probability one. Conditional on this assignment, the manager's payoff is $V(hi)=(1-\gd)\underline v+\gd[qV(hiS)+(1-q)V(hiF)]\leq(1-\gd)\underline v+\gd\overline v\leq0$. Because this assignment is in the support of the manager's strategy and she can guarantee zero, its conditional payoff must be nonnegative. Thus both inequalities must bind. If $(1-\gd)\underline v+\gd\overline v<0$, this is an immediate contradiction. If $(1-\gd)\underline v+\gd\overline v=0$, then $q\in(0,1)$ implies $V(hiS)=V(hiF)=\overline v$. Attaining $\overline v$ requires an effortful assignment in every subsequent period on path. Either continuation is therefore a no-shirking equilibrium attaining the manager's first-best, contradicting $c>\overline c_n$ by \cref{thm:secondbest}. Consequently, every on-path assignee exerts effort with positive probability.

If the task is never assigned on path, the manager's payoff is zero and cannot exceed her no-shirking optimum. Suppose henceforth that some assignment occurs on path. We next show that every worker's continuation payoff lies in $[0,r]$. Nonnegativity follows because all worker stage payoffs are nonnegative. Let $M:=\max_{i\in N}\sup_{h^*} U^i(h^*)$, which is finite because stage payoffs are bounded. When worker $i$ is assigned, effort belongs to the support of his equilibrium strategy and is therefore a best response. His equilibrium payoff consequently equals his payoff from choosing effort: $U^i(hi)=\gd[pU^i(hiS)+(1-p)U^i(hiF)]\leq\gd M$. When worker $i$ is unassigned, his current payoff is $r$, so his continuation payoff is at most $(1-\gd)r+\gd M$. Taking the supremum gives $M\leq(1-\gd)r+\gd M$, and hence $M\leq r$.

For each worker $i$, define $\widetilde x_i(h):=1-U^i(h)/r$. Thus $0\leq\widetilde x_i(h)\leq1$. These quantities coincide with assignment frequencies in a no-shirking equilibrium; we refer to them as effective assignment frequencies and give their frequency representation below. Because effort is a best response after every on-path assignment, the assignee's promise-keeping and incentive constraints imply
\begin{align}
\widetilde x_i(hi)
&=
(1-\gd)+\gd\left[p\widetilde x_i(hiS)+(1-p)\widetilde x_i(hiF)\right],
\label{eq:oa-unrestricted-promise}\\
\widetilde x_i(hiF)-\widetilde x_i(hiS)
&\geq c.
\label{eq:oa-unrestricted-ic}
\end{align}
The first equality remains valid when the assignee mixes: because effort is in the support of his strategy, choosing effort gives exactly his equilibrium payoff. The second inequality binds whenever he also shirks with positive probability.

Let $\widetilde B:=\max_{i\in N}\sup_{h^*}\widetilde x_i(h^*)$. Since an on-path assignee has $U^i(hi)\leq\gd r$, his effective assignment frequency is at least $1-\gd$, so $\widetilde B>0$. Fix a worker $i$ attaining the maximum over $N$ and choose a sequence of post-decision histories along which his effective assignment frequency converges to $\widetilde B$. Whenever he is unassigned, his current payoff is $r$, so his effective assignment frequency is at most $\gd\widetilde B<\widetilde B$. He must therefore be assigned along all sufficiently late elements of the sequence. At each such history, \eqref{eq:oa-unrestricted-ic} and $\widetilde x_i(hiF)\leq\widetilde B$ imply $\widetilde x_i(hiS)\leq\widetilde B-c$. Substituting these bounds into \eqref{eq:oa-unrestricted-promise} and taking the limit gives $\widetilde B\leq(1-\gd)+\gd[p(\widetilde B-c)+(1-p)\widetilde B]$. Moreover, nonnegativity and \eqref{eq:oa-unrestricted-ic} imply $\widetilde B\geq c$. Consequently,
\begin{align}
c\leq\widetilde B\leq1-\frac{\gb}{1-\gb}c.
\label{eq:oa-unrestricted-individual-bound}
\end{align}
These inequalities require $c\leq1-\gb$. In particular, if $c>1-\gb$, no assignment can occur on path in any equilibrium satisfying the stated manager-payoff condition, and permanent vacancy is optimal. When assignment does occur, $c\leq1-\gb$ also implies $s/r\leq\gb(p-q)/p<1$. Thus $s<r$ follows from equilibrium incentives rather than being an additional assumption.

For every worker $i$, let 
$$e_i(h):=(1-\gd)\bE\!\left[\sum_{t=0}^{\infty}\gd^t\mathbf 1\{i\textnormal{ is assigned and exerts effort }t \textnormal{ periods after } h\} \middle|h \right]$$
and 
$$
z_i(h):=(1-\gd)\bE\!\left[\sum_{t=0}^{\infty}\gd^t\mathbf 1\{i\textnormal{ is assigned and shirks }t \textnormal{ periods after } h\} \middle| h \right]$$
denote his average frequencies of effortful and shirking assignments. His payoff satisfies $U^i(h)=r[1-e_i(h)-z_i(h)]+sz_i(h)$, so $\widetilde x_i(h)=e_i(h)+(1-s/r)z_i(h)$. Since $s<r$ and at most one worker is assigned in each period, the aggregate effective assignment frequency $\widetilde A(h):=\sum_{i=1}^n\widetilde x_i(h)$ satisfies $0\leq\widetilde A(h)\leq1$. The manager's payoff satisfies
\begin{align}
V(h)
=
\overline v\sum_{i=1}^n e_i(h)+\underline v\sum_{i=1}^n z_i(h) \le
\overline v\sum_{i=1}^n e_i(h)
\leq
\overline v\widetilde A(h).
\label{eq:oa-unrestricted-manager-bound}
\end{align}

It remains to bound $\widetilde A(h)$. For $k=1,\ldots,n$, define
$$\widetilde H_k:=\max_{\substack{C\subseteq N\\|C|=k}}\sup_{h^*}\sum_{j\in C}\widetilde x_j(h^*),$$
and set $\widetilde H_0:=0$. Nonnegativity of effective assignment frequency implies $\widetilde H_k\geq\widetilde B>0$, while the definitions imply $\widetilde H_k\leq\widetilde H_{k-1}+\widetilde B$.

Fix a group $C$ attaining the maximum over groups of $k$ workers and choose a sequence of post-decision histories along which its total effective assignment frequency converges to $\widetilde H_k$. If no member of $C$ is currently assigned, the group's effective assignment frequency is at most $\gd\widetilde H_k<\widetilde H_k$. Thus the current assignee eventually belongs to $C$. Because $C$ is finite, restrict attention to a subsequence along which the same worker $i\in C$ is assigned. Write these histories as $h_mi$, let $\alpha_m\in(0,1]$ be his probability of effort, and let $p_m:=q+\alpha_m(p-q)$ be the resulting success probability. Denote the group's effective assignment frequency at $h_mi$ by $g_m$. Following success, \eqref{eq:oa-unrestricted-ic} bounds the assignee's effective assignment frequency by $\widetilde B-c$, while the other $k-1$ workers' total effective assignment frequency is at most $\widetilde H_{k-1}$. Following failure, the group's total is at most $\widetilde H_k$. Its current expected effective assignment frequency is $1-(1-\alpha_m)s/r$. Aggregate promise keeping therefore gives
\begin{align}
g_m
&\le
(1-\gd)\left( 1-(1-\alpha_m)\frac{s}{r}\right) +
\gd\left( p_m(\widetilde H_{k-1}+\widetilde B-c)+(1-p_m)\widetilde H_k\right).
\label{eq:oa-unrestricted-group-bound}
\end{align}
Because $p-p_m=(1-\alpha_m)(p-q)$ and $(1-\gd)(1-\alpha_m)s/r=\gd(p-p_m)c$, the right-hand side of \eqref{eq:oa-unrestricted-group-bound} equals $(1-\gd)+\gd[p(\widetilde H_{k-1}+\widetilde B-c)+(1-p)\widetilde H_k]+\gd(p-p_m)(\widetilde H_k-\widetilde H_{k-1}-\widetilde B)$. The last term is nonpositive because $\widetilde H_k\leq\widetilde H_{k-1}+\widetilde B$. Taking the limit along the subsequence yields $\widetilde H_k\leq(1-\gd)+\gd[p(\widetilde H_{k-1}+\widetilde B-c)+(1-p)\widetilde H_k]$. Rearranging and applying \eqref{eq:oa-unrestricted-individual-bound} gives $\widetilde H_k\leq1-\frac{\gb}{1-\gb}c+\gb\widetilde H_{k-1}$.

Iterating this recursion from $\widetilde H_0=0$ yields $\widetilde H_n\leq(1-\gb^n)[(1-\gb)-\gb c]/(1-\gb)^2$. Combining $\widetilde A(h)\leq\widetilde H_n$, $\widetilde A(h)\leq1$, and \eqref{eq:oa-unrestricted-manager-bound}, we obtain
\begin{align}
V(h)
&\leq
\overline v\min\left(
1,
\frac{(1-\gb^n)\left[(1-\gb)-\gb c\right]}{(1-\gb)^2}
\right).
\label{eq:oa-unrestricted-payoff-bound}
\end{align}
For $c\leq1-\gb$, \cref{thm:secondbest} provides a no-shirking equilibrium attaining the right-hand side of \eqref{eq:oa-unrestricted-payoff-bound}. For $c>1-\gb$, we have shown that every equilibrium leaves the task vacant on path. The no-shirking equilibrium characterized in \cref{thm:secondbest} therefore attains the manager's maximum payoff over all equilibria.
\end{proof}

\subsection{Desirable task}
\label{sec:desirable-task}

In this Appendix, we consider the desirable-task variant of the model. Define
\begin{align}
\widehat c_n
:=
(1-\widehat\beta)
\frac{
1-\widehat\beta^{n-1}
}{
1-\widehat\beta^n
},
\label{eq:desirable-definitions}
\end{align}
where $\widehat\beta$ is given in \eqref{eq:hatgb}. The following analogue of \cref{thm:secondbest} holds. Its proof mirrors that of \cref{thm:secondbest}, with $\widehat\beta$ in place of $\beta$ and the roles of success and failure interchanged.

\begin{prop*}[Desirable-task optimum]
\label{prop:desirable-task}
If $c\leq1-\widehat\beta$, the manager's optimal payoff is
\begin{align}
\begin{cases}
\overline v,
&
c\leq\widehat c_n,
\\[6pt]
\displaystyle
\overline v
\frac{
(1-\widehat\beta^n)
((1-\widehat\beta)-\widehat\beta c)}{
(1-\widehat\beta)^2
},
&
\widehat c_n<c\leq1-\widehat\beta.
\end{cases}
\label{eq:desirable-payoff}
\end{align}
This payoff is attained by a no-shirking equilibrium with a reverse first-order rotation rule. Under this rule, success retains the current assignee, whereas failure moves him to the back of the rotation and advances the next worker to the task. A rotation phase ends after every worker has failed once. The vacancy-phase length $L$ at the end of each rotation phase is selected as follows.
\begin{enumerate}\itemsep0em
\item[1.] If $c\leq\widehat c_n$, set $L=0$.
\item[2.] If $\widehat c_n<c<1-\widehat\beta$, draw
\begin{align}
L
=
\begin{cases}
\ell,
&
\textnormal{with probability }
\displaystyle
\frac{
((1-\widehat\beta)-c)/(
\widehat\beta^{n-1}
(
(1-\widehat\beta)-\widehat\beta c
))
-\gd^{\ell+1}
}{
\gd^\ell-\gd^{\ell+1}
},
\\[0.5em]
\ell+1,
&
\textnormal{otherwise},
\end{cases}
\label{eq:desirable-vacancy-lottery}
\end{align}
where $\ell\in \bN$ is any nonnegative integer satisfying
\begin{align}
\gd^{\ell+1}
\leq
\frac{
(1-\widehat\beta)-c
}{
\widehat\beta^{n-1}
\left[
(1-\widehat\beta)-\widehat\beta c
\right]
}
\leq
\gd^\ell.
\label{eq:desirable-optimal-support}
\end{align}
\item[3.] If $c=1-\widehat\beta$, set $L=\infty$.
\end{enumerate}
Finally, if $c>1-\widehat\beta$, the manager's optimal payoff is zero, attained by permanent vacancy.
\end{prop*}

\begin{proof}[Proof of \cref{prop:desirable-task}]Fix an arbitrary no-shirking equilibrium. We adopt the notation from the proof of \cref{thm:secondbest}. Suppose worker $i$ is assigned after history $h$. His average assignment frequency satisfies the same promise-keeping equation \eqref{eq:sb-promise-keeping}. His continuation-payoff difference between success and failure is $r[x_i(hiS)-x_i(hiF)]$. Therefore, his effort incentive constraint is
\begin{align}
x_i(hiS)-x_i(hiF)
\geq
c.
\label{eq:desirable-ic}
\end{align}
Define $B:=\max_{i\in N}\sup_{h^*} x_i(h^*)$. If $B=0$, no worker is assigned on path. Suppose that $B>0$. The same maximizing-sequence argument used in the proof of \cref{lem:individual-assignment-frequency} implies that a sequence approaching $B$ can be chosen along which the same worker $i$ is currently assigned. At each such history, $x_i(hiS)\leq B$, so \eqref{eq:desirable-ic} implies $x_i(hiF)\leq B-c$. Substituting these bounds into \eqref{eq:sb-promise-keeping} and taking the limit gives $B\leq(1-\gd)+\gd[pB+(1-p)(B-c)]$. Since $\widehat\beta=\gd(1-p)/(1-\gd p)$ and $1-\widehat\beta=(1-\gd)/(1-\gd p)$, rearranging yields
\begin{align}
B
\leq
1-
\frac{
\widehat\beta
}{1-\widehat\beta
}c =: \overline B^D(c).
\label{eq:desirable-individual-bound}
\end{align}
An effortful assignment also requires $B\geq c$: by \eqref{eq:desirable-ic} and nonnegativity, $x_i(hiS)\geq c$, while $x_i(hiS)\leq B$. Combining this inequality with \eqref{eq:desirable-individual-bound} gives $c\leq1-\widehat\beta$. Consequently, if $c>1-\widehat\beta$, no effortful assignment can occur on path in a no-shirking equilibrium.

Suppose henceforth that $c\leq1-\widehat\beta$. Under this restriction, $\overline B^D(c)>0$, so \eqref{eq:desirable-individual-bound} also holds when $B=0$. For $k=1,\ldots,n$, define $H_k:=\max_{\substack{C\subseteq N\\|C|=k}}\sup_{h^*}\sum_{j\in C}x_j(h^*)$ and set $H_0:=0$. If $H_k=0$, the recursion below is immediate. Suppose that $H_k>0$. The same maximizing-sequence argument used in the proof of \cref{lem:group-assignment-frequency} implies that a sequence approaching $H_k$ can be chosen along which the same group $C$ contains the current assignee $i$. Following success, the group's total assignment frequency is at most $H_k$. Following failure, \eqref{eq:desirable-ic} and $x_i(hiS)\leq B$ imply $x_i(hiF)\leq B-c$, while the remaining $k-1$ workers have total average assignment frequency at most $H_{k-1}$. Aggregate promise keeping thus gives $H_k\leq(1-\gd)+\gd[pH_k+(1-p)(H_{k-1}+B-c)]$. Rearranging yields $H_k\leq(1-\widehat\beta)+\widehat\beta(H_{k-1}+B-c)$. Applying \eqref{eq:desirable-individual-bound} and using $(1-\widehat\beta)+\widehat\beta[\overline B^D(c)-c]=\overline B^D(c)$ yields
\begin{align}
H_k
\leq
\overline B^D(c)
+
\widehat\beta H_{k-1}.
\label{eq:desirable-recursion}
\end{align}
This is the counterpart of the average assignment-frequency recursion in \cref{lem:group-assignment-frequency}, with $\gb$ replaced by $\widehat\beta$ and the roles of success and failure interchanged.

Iterating \eqref{eq:desirable-recursion} from $H_0=0$ gives
\begin{align*}
H_n
\le
\overline B^D(c)
\sum_{j=0}^{n-1}\widehat\beta^j =
\frac{
(1-\widehat\beta^n)
((1-\widehat\beta)-\widehat\beta c)
}{
(1-\widehat\beta)^2
}.
\end{align*}
Because $A(h^*)\leq1$ and $A(h^*)\leq H_n$ at every post-decision on-path continuation, while every pre-decision average assignment frequency is a finite convex combination of post-decision average assignment frequencies, the manager's continuation payoff at every on-path continuation is bounded above by $\overline v$ times
\begin{align}
\min\left(
1,
\frac{
(1-\widehat\beta^n)
((1-\widehat\beta)-\widehat\beta c)
}{
(1-\widehat\beta)^2
}
\right).
\label{eq:desirable-upper-bound}
\end{align}
The second term in \eqref{eq:desirable-upper-bound} is at least one if and only if $c\leq(1-\widehat\beta)(1-\widehat\beta^{n-1})/(1-\widehat\beta^n)=\widehat c_n$.

It remains to verify that the stated reverse first-order rotation rule attains this bound in a no-shirking equilibrium. The verification follows the proof of \cref{thm:secondbest}, with $\gb$ replaced by $\widehat\beta$ and success and failure interchanged. We omit the details.
\end{proof}

\end{appendices}

\pagebreak

{
\bibliographystyle{dcu}
\bibliography{references}

@article{mylovanov2008task,
  title={{Task Scheduling and Moral Hazard}},
  author={Mylovanov, Tymofiy and Schmitz, Patrick W},
  journal={Economic Theory},
  volume={37},
  number={2},
  pages={307--320},
  year={2008},
  publisher={Springer}
}

@article{demarzo2006optimal,
  title={{Optimal Security Design and Dynamic Capital Structure in a Continuous-Time Agency Model}},
  author={DeMarzo, Peter M and Sannikov, Yuliy},
  journal={The Journal of Finance},
  volume={61},
  number={6},
  pages={2681--2724},
  year={2006},
  publisher={Wiley Online Library}
}

@article{butler2017motivating,
  title={{Motivating Workers through Task Assignment: A Dynamic Model of Up-and-Down Competition for Status}},
  author={Butler, Julianna M and Gilpatric, Scott M and Vossler, Christian A},
  year={2017},
  journal={Working paper, University of Tennessee}
}

@article{watt2026paying,
  title={{Paying With Promises}},
  author={Watt, Mitchell},
  year={2026},
  journal={Working paper, Monash Business School}
}

@article{rogerson1985first,
  title={{The First-Order Approach to Principal-Agent Problems}},
  author={Rogerson, William P},
  journal={Econometrica},
  pages={1357--1367},
volume={53},
number={6},  
  year={1985},
  publisher={JSTOR}
}

@article{ely2005belief,
title={{Belief-Free Equilibria In Repeated Games}},
author={Ely, Jeffrey C and H{\"o}rner, Johannes and Olszewski, Wojciech},
journal={Econometrica},
volume={73},
number={2},
pages={377--415},
year={2005},
publisher={Wiley Online Library}
}

@incollection{garicano2013hierarchies,
author    = {Garicano, Luis and Van Zandt, Timothy},
title     = {{Hierarchies and the Division of Labor}},
booktitle = {The Handbook of Organizational Economics},
editor    = {Gibbons, Robert and Roberts, John},
publisher = {Princeton University Press},
year      = {2013},
pages     = {604--654}
}

@article{taylor1995digging,
title={{Digging for Golden Carrots: An Analysis of Research Tournaments}},
author={Taylor, Curtis R},
journal={The American Economic Review},
pages={872--890},
volume={85},
number={4},
year={1995},
publisher={JSTOR}
}

@article{nalebuff1983prizes,
title={{Prizes and Incentives: Towards A General Theory of Compensation and Competition}},
author={Nalebuff, Barry J and Stiglitz, Joseph E},
journal={The Bell Journal of Economics},
pages={21--43},
volume={14},
number={1},
year={1983},
publisher={JSTOR}
}

@article{halac2021rank,
title={{Rank Uncertainty In Organizations}},
author={Halac, Marina and Lipnowski, Elliot and Rappoport, Daniel},
journal={American Economic Review},
volume={111},
number={3},
pages={757--786},
year={2021},
publisher={American Economic Association 2014 Broadway, Suite 305, Nashville, TN 37203}
}

@article{eliaz2026clerks,
title={{Clerks}},
author={Eliaz, Kfir and Fershtman, Daniel and Frug, Alexander},
year={2026},
journal={Working paper, Tel Aviv University\!}
}

@article{bird2019dynamic,
title={{Dynamic Non-Monetary Incentives}},
author={Bird, Daniel and Frug, Alexander},
journal={American Economic Journal: Microeconomics},
volume={11},
number={4},
pages={111--150},
year={2019},
publisher={American Economic Association 2014 Broadway, Suite 305, Nashville, TN 37203-2425}
}

@article{georgiadis2015projects,
title={{Projects and Team Dynamics}},
author={Georgiadis, George},
journal={The Review of Economic Studies},
volume={82},
number={1},
pages={187--218},
year={2015},
publisher={Oxford University Press}
}

@article{bonatti2011collaborating,
title={{Collaborating}},
author={Bonatti, Alessandro and H{\"o}rner, Johannes},
journal={American Economic Review},
volume={101},
number={2},
pages={632--663},
year={2011},
publisher={American Economic Association}
}

@article{rayo2007relational,
title={{Relational Incentives and Moral Hazard In Teams}},
author={Rayo, Luis},
journal={The Review of Economic Studies},
volume={74},
number={3},
pages={937--963},
year={2007},
publisher={Wiley-Blackwell}
}

@article{guo2020dynamic,
title={{Dynamic Allocation Without Money}},
author={Guo, Yingni and H{\"o}rner, Johannes},
journal={Working paper, Toulouse School of Economics\!},
year={2020}
}

@article{che2001optimal,
title={{Optimal Incentives For Teams}},
author={Che, Yeon-Koo and Yoo, Seung-Weon},
journal={American Economic Review},
volume={91},
number={3},
pages={525--541},
year={2001},
publisher={American Economic Association}
}

@article{winter2004incentives,
title={{Incentives and Discrimination}},
author={Winter, Eyal},
journal={American Economic Review},
volume={94},
number={3},
pages={764--773},
year={2004},
publisher={American Economic Association}
}

@article{athey2001optimal,
title={{Optimal Collusion with Private Information}},
author={Athey, Susan and Bagwell, Kyle},
journal={RAND Journal of Economics},
pages={428--465},
volume={32},
number={3},
year={2001},
publisher={JSTOR}
}

@article{andrews2016allocation,
title={{The Allocation of Future Business: Dynamic Relational Contracts with Multiple Agents}},
author={Andrews, Isaiah and Barron, Daniel},
journal={American Economic Review},
volume={106},
number={9},
pages={2742--2759},
year={2016},
publisher={American Economic Association 2014 Broadway, Suite 305, Nashville, TN 37203}
}

@article{de2021selecting,
title={{On Selecting The Right Agent}},
author={De Clippel, Geoffroy and Eliaz, Kfir and Fershtman, Daniel and Rozen, Kareen},
journal={Theoretical Economics},
volume={16},
number={2},
pages={381--402},
year={2021},
publisher={Wiley Online Library}
}

@article{abreu1990toward,
title={{Toward a Theory of Discounted Repeated Games with Imperfect Monitoring}},
author={Abreu, Dilip and Pearce, David and Stacchetti, Ennio},
journal={Econometrica},
volume={58},
number={5},
pages={1041--1063},
year={1990},
publisher={JSTOR}
}

@article{fudenberg1994folk,
title={{The Folk Theorem with Imperfect Public Information}},
author={Fudenberg, Drew and Levine, David K and Maskin, Eric},
journal={Econometrica},
volume={62},
number={5},
pages={997--1039},
year={1994},
}

@article{lipnowski2020repeated,
title={{Repeated Delegation}},
author={Lipnowski, Elliot and Ramos, Joao},
journal={Journal of Economic Theory},
volume={188},
pages={105040},
year={2020},
publisher={Elsevier}
}

@article{li2017power,
title={{Power Dynamics in Organizations}},
author={Li, Jin and Matouschek, Niko and Powell, Michael},
journal={American Economic Journal: Microeconomics},
volume={9},
number={1},
pages={217--241},
year={2017},
publisher={American Economic Association 2014 Broadway, Suite 305, Nashville, TN 37203-2425}
}

@article{board2011relational,
title={{Relational Contracts and the Value of Loyalty}},
author={Board, Simon},
journal={American Economic Review},
volume={101},
number={7},
pages={3349--3367},
year={2011},
publisher={American Economic Association}
}

@article{fudenberg1994efficiency,
title={{Efficiency and Observability with Long-Run and Short-Run Players}},
author={Fudenberg, Drew and Levine, David K},
journal={Journal of Economic Theory},
volume={62},
number={1},
pages={103--135},
year={1994},
publisher={Elsevier}
}

@article{mailath2001,
title={Who {W}ants a {G}ood {R}eputation?},
author={Mailath, George and Samuelson, Larry},
journal={The Review of Economic Studies},
volume={68},
number={2},
pages={415--441},
year={2001},
publisher={Wiley-Blackwell}
}
}




\end{document}